\documentclass[iop,apj]{emulateapj}
\usepackage{amsmath,amssymb,amstext,comment}

\usepackage[breaklinks,colorlinks,citecolor=blue,linkcolor=magenta]{hyperref}

\newcommand{\pd}[2]{\frac{\partial #1}{\partial #2}}

\usepackage[all]{hypcap} 
\usepackage{aas_macros}
\usepackage{natbib}
\shorttitle{Numerical and Analytical TTV}
\shortauthors{Hadden \& Lithwick}

\begin{document}

\title{ 
 	    Numerical and Analytical Modelling of Transit Time Variations
} 
\author{Sam Hadden,Yoram Lithwick}
\affil{Department of Physics \& Astronomy, Northwestern University, Evanston, IL 60208, USA \& Center for Interdisciplinary Exploration and Research in Astrophysics (CIERA)}

\begin{abstract}
	  We develop and apply methods to extract planet masses and eccentricities from observed transit time variations (TTVs).   
	First, we derive simple analytic expressions for the TTV that include the effects of both first- and second-order resonances. 
	Second, we use N-body Markov chain Monte Carlo (MCMC) simulations, as well as the analytic formulae, to measure the masses and eccentricities of ten planets discovered by {\it Kepler} that 	have not previously been analyzed.   Most of the ten planets have low densities. Using the analytic expressions to partially circumvent degeneracies, we measure small eccentricities of a few percent or less.	
\end{abstract}

\maketitle

\section{Introduction}
\label{sec:intro}	
The {\it Kepler} mission has revealed a wide diversity of extrasolar planetary systems. Super-Earth and sub-Neptune planets with radii in the range of $\sim$1--4 $R_\oplus$ have been shown to be abundant in the Galaxy, even though no such planet exists in our own Solar System. 
Determining the compositions of these abundant planets is important for understanding the planet formation process. 
 The orbital architectures of many of {\it Kepler}'s multiplanet system are starkly different from our Solar System's as well. 
 Precise measurements of the dynamical states of multi-planet systems offer important clues about their origins and evolution.
	
Transit timing variations (TTVs) are a powerful tool for measuring masses and eccentricities in multi-planet systems \citep{2005MNRAS.359..567A,Holman:2005jf}. Planets near mean-motion resonances (MMRs) often exhibit
 large TTV signals, allowing for sensitive measurements of the properties of low-mass planets.  However the conversion of a TTV signal to planet properties is often plagued by a degeneracy between planet masses and eccentricities \citep[][hereafter LXW]{2012ApJ...761..122L}. Nonetheless, N-body analyses have provided a number of TTV systems in which planet masses are apparently well constrained and not subject to the predicted degeneracies \citep[e.g.,][]{2015Natur.522..321J,2014ApJ...795..167S,2014ApJ...785...15J,2013ApJ...770..131L,Nesvorny:2013cb,Huber2013misalign,2013ApJ...778..185M,2012Natur.487..449S,Carter:2012gq,Cochran:2011kg,kipping2014ttv}.  \citet{2014ApJ...790...58N} and  \citet{Deck:2014vr}   show that the the mass-eccentricity degeneracy can be broken provided that the effects of the planets' successive conjunctions  are seen with sufficient signal to noise in their TTVs.
    
    Characterizing multi-planet systems on the basis of transit time observations  involves fitting noisy data in a high-dimensional parameter space. Bayesian parameter estimation via Markov-Chain Monte Carlo (MCMC) is well suited to handle such problems and has been applied to the analysis of TTVs previously by numerous authors \citep[e.g.][]{2015Natur.522..321J,2014ApJ...795..167S,2013ApJ...778..185M,Huber2013misalign,2012Natur.487..449S}. (Also see \citet{kipping2014ttv} for an alternative Bayesian approach to dynamical modeling of TTVs.)
	
    In this paper we derive  analytic formulae for the TTVs of planets near first- and second-order MMRs.\footnote{
       \citet{Deck:2015tx} also derive  analytic formulae for TTVs near first- and second-order MMRs.  Their paper was posted to
    arxiv.org  shortly before this one. 
  	}
    We conduct MCMC analyses using both the analytic model and N-body integrations to infer masses and eccentricities of ten planets in four {\it Kepler} multi-planet systems.
    The analytic model elucidates the degeneracies inherent to inverting TTVs and provides a complimentary approach to parameter inference.

 The paper is organized as follows: We  summarize the analytic TTV model in Section \ref{sec:analytic_model}.
 In Section \ref{sec:systems} we detail our methods for inverting TTVs using both N-body and analytic methods 
  and apply them to four {\it Kepler} multi-planet systems. 
 We summarize our results and conclude in Section \ref{sec:summary}.

\section{analytic TTV}
\label{sec:analytic_model}

\subsection{The Formulae}
\label{sec:formulae}
	In Appendix \ref{sec:appendix} we derive the analytic TTV  for a pair of  planets that lie
	near (but not in) either a first order  ($J$:$J$-1) mean motion resonance (MMR)
	or a second order one ($K$:$K$-2).\footnote{ Throughout this paper capital `$J$'  refers to the nearest  first order MMR and `$K$'  to the nearest second order MMR (with $J$ and $K$ both positive),  while lowercase $j$'s refers to generic MMR's.
		The distinction  between  $j$ and $J$ is necessary because many different MMR's contribute to the TTV of a pair of planets, not just the closest one.
		}
	These formulae should describe  the vast majority of TTV's observed by {\it Kepler}.\footnote{
	The formulae are invalid for planets that are either librating in resonance, near
	third or higher  order MMR, or highly eccentric or inclined. In all cases we have examined, TTVs in systems with three or more planets are well approximated as sums over  pairwise interactions.
	}

    Here we provide a qualitative overview of the  formulae because it will help in understanding
  how well  masses and eccentricities can be inferred 
   from observed TTV's (Section \ref{sec:systems}). 
     We focus  first on the case 
  of a planet perturbed by an exterior  companion  near its $J$:$J$-1 resonance;  we then address the other cases of interest, which are almost trivial extensions.
The planet's TTV is
 $\delta t\equiv O-C$, where $O$ is 
  its observed time of transit and $C$ is 
the time calculated from  its average orbital period, under the assumption of a perfectly periodic orbit. 
To derive a simple expression for $\delta t$, we expand in powers of the planets' eccentricities, which is appropriate because  
{\it Kepler} planets typically have 
  $e\lesssim 0.2$  (e.g., see  \citet{2014ApJ...787...80H} and below).
However, the most important contributions are not necessarily zeroth order in $e$.
That is because there is a second  small parameter that can compensate for a small  $e$: the fractional distance to the nearest
first-order MMR:
			\begin{equation}
			\Delta = \frac{P'}{P}\frac{J-1}{J}-1 \label{eq:delta_defn}
			\end{equation}
	where $P$ and $P'$ are the orbital periods of the inner and outer planet.  
	Planets near resonance ($|\Delta|\ll 1$) have particularly large TTVs because
	the gravitational perturbations add coherently over many orbital periods. Observationally, 
   	{\it Kepler} pairs with detected TTV's typically have $|\Delta|\sim 1-5\%$.

After expanding in $e$, we reshuffle terms to express the TTV as a sum of three terms
 that differ in their frequency dependence: 
\begin{eqnarray}
\delta t = \delta t_{\cal F}+\delta t_{\cal C} +\delta t_{\cal S} \label{eq:decomp} \ ,
\end{eqnarray}
where $\delta t_{\cal F}$ and $\delta t_{\cal S}$ are sinusoidal (with different frequencies) and $\delta t_{\cal C}$ is the sum of many sinusoidal terms. The three components are given explicitly in 
(Eqs. \ref{eq:tfund}--\ref{eq:tsec})
and are described  in the following.

	\begin{itemize}
		\item  $\delta t_{\cal F}$: The ``fundamental'' (or alternatively  ``first harmonic'')
		has the longest period and typically has the
		   largest amplitude (LXW).
		Its  period is that of the planets' line of conjunction (the ``superperiod'')
			\begin{eqnarray}
	     		P_{\rm super}=\left| {J\over P'}-{J-1\over P} \right|^{-1}=  {P'\over J|\Delta|} \ , 
			\label{eq:super}
			\end{eqnarray}
		and its amplitude is, within order-unity constants\footnote{ 
		The ``order-unity constants''  that are dropped from the TTV expressions in this section depend only  on the planets' period ratio  
		and
		 the MMR integer indices.}

		 	\begin{eqnarray}
				\delta t_{\cal F}\sim  {\mu' P\over 2\pi|\Delta|}\cdot {\rm max}\left\{1,{|{\cal Z}|\over|\Delta|}	\right\} \label{eq:t_F_amp}
			\end{eqnarray}
			where $\mu'$ is the ratio of the outer planet's mass to the star's mass, and 
			\begin{eqnarray}
			{\cal Z} &\equiv& \frac{f_{27}^{J} z + f_{31}^{J} z'}{\sqrt{(f_{27}^{J})^2+ (f_{31}^{J})^2}} \label{eq:Z_def}\\
			&\approx&  
			{z'-z\over \sqrt{2}} \ . \label{eq:zapprox}
			\end{eqnarray}
			is an important variable that 
			consolidates the effect of the planets' eccentricities on the TTV;\footnote{
			 LXW employ the variable
			$Z_{\rm free}$ rather than ${\cal Z}$, which differs in its normalization.
			  We prefer here  ${\cal Z}$, because it approximately satisfies   Eq. \ref{eq:zapprox}.   }
			in the above definition, 
			$z$ is the complex eccentricity of the inner planet  ($z\equiv ee^{i\varpi}$), 
			$z'$ of the outer, 
			and
			the $f$'s  are Laplace coefficients (in
			 the notation of  \citet{1999ssd..book.....M}). Numerical values for the $f$'s  are
			  tabulated 
			 in the Appendix of LXW.  
			The approximate form of Eq. \ref{eq:zapprox}---which is independent of $J$---is valid to within 
			around 10\% fractional error in the coefficients of the $z$'s (for $J> 2$).

	\item $\delta t_{\cal C}$: The ``chopping'' TTV is a sum of many sinusoids that have higher frequencies
	than the fundamental. These were first derived by \citet{Deck:2014vr}. 
	  The amplitude of each of the terms is
		\begin{eqnarray}
		 \delta t_{\cal C}\sim {\mu'P\over 2\pi} \left({\frac{P}{P-P'}}\right)^2\ ,
		\end{eqnarray}
		within order unity constants.  
		All first order and zeroth order MMR's contribute with roughly this same amplitude---except
		for the nearby $J$:$J$-1, whose  contribution produces $\delta t_{\cal F}$.  
		The chopping is independent of eccentricity because there are no resonant denominators 		(i.e., factors of $1/|\Delta|$), and hence the zeroth order term in the eccentricity expansion
		is adequate. 
		Physically, chopping is caused by  kicks at conjunctions which can suddenly change the orbit.  	
		As a result, the TTV exhibits a strong ``chopping'' spike at each transit that follows a conjunction \citep{2014ApJ...790...58N,Deck:2014vr}	.

	\item  $\delta t_{\cal S}$: The ``secondary''  (or alternatively  ``second harmonic'' or ``second order MMR'')  term has
	twice the frequency of $\delta t_{\cal F}$.  It is caused by proximity to a second order MMR:
	i.e., the $2J$:$2J$-2 MMR for a planet pair near the $J$:$J$-1.  We derive its effect in the Appendix.  Its amplitude is, within order unity constants, a factor of $|{\cal Z}|$ 
	smaller than the fundamental:
	\begin{eqnarray}
				\delta t_{\cal S}\sim  {\mu' P \over2\pi|\Delta|}|{\cal Z}| \cdot {\rm max}\left\{1,{|{\cal Z}|\over|\Delta|}	\right\}
	\end{eqnarray}
	\end{itemize}
 
 Having completed the discussion of an interior planet's TTV near a first order resonance, we turn
now to the other cases of interest.  First, for an exterior planet near a first order resonance, the
discussion above carries through unchanged, after 
replacing mass and period appropriately, i.e.  $\mu'\rightarrow\mu$ and  $P\rightarrow P'$. 
The only other difference is in the order-unity coefficients, which are in any case dropped above.  The full formulae that include the order-unity coefficients are given in  \ref{sec:outerPlanet}.
Second, for planets near a second-order $K:K$-2 resonance (with $K$ an odd number), the 
inner planet's TTV is $\delta t=\delta t_{\cal C}+\delta t_{\cal S}$, i.e., there is no 
  $\delta t_{\cal F}$ because it may now be included with the other chopping terms.  The secondary TTV is unchanged, with $2J\rightarrow K$.  See  \ref{sec:expand}--\ref
  {sec:outerPlanet} for the full formulae.

 \subsection{Inferring Planet Parameters}
\label{sec:inferring}
One approach
to inferring planet parameters from  TTV
 is  with  MCMC simulations (Section \ref{sec:methods}).
 But to understand the MCMC results and to evaluate, for example, the effects
of  degeneracies and assumed priors on those results, we develop in this section a complementary approach, based 
on the analytic formulae. 

For definiteness, we focus here on a two planet system 
near a $J$:$J$-1 MMR.
Each planet has, essentially, three unknown parameters: its mass, eccentricity, and longitude
of periapse  (or equivalently $\mu$ and  complex $z$).
  It also has two additional parameters that are simple to determine accurately, and hence we consider ``known'': its semimajor axis
and mean longitude at epoch (or equivalently
period and the time of a particular transit). 
For completeness, we note that there are two additional parameters  per planet associated with inclinations, but we ignore
them here as they usually have a lesser effect on TTV's (see Appendix \ref{sec:inclinations}).

To clarify the parameter inference problem, we rewrite the inner planet's TTV  (Eqs. \ref{eq:tfund}--\ref{eq:tsec}) in a form that
highlights the  unknowns ($\mu'$ and ${\cal Z}$):

\begin{eqnarray}
\delta t_{\cal F}&=& \mu' 
\left(A+B{\cal Z^*}\right)ie^{iJ\lambda'}  +c.c.
\label{eq:ttv1X}
\\
\delta t_{\cal C}&=&
\mu'  \sum_{j>0,j\ne J} C_j ie^{ij\lambda'}  +c.c. \label{eq:ttv2X}
\\
\delta t_{\cal S} &=& \mu' \left(  D{\cal Z^*}+E{\cal Z}^{*2}\right)ie^{i2J\lambda'}
+c.c.
\label{eq:ttv3X}
\end{eqnarray}
where ``c.c.'' denotes the complex conjugate of the preceding term and
 the  coefficients $A$ through $E$  are real-valued
 ``known'' numbers, i.e., determined by the planets' periods. 
   In addition,  
$\lambda'={\rm const}+{t}(2\pi/P')$ is the mean longitude of the outer planet. Note that the period of $\delta t_{\cal F}$
is the superperiod (Eq. \ref{eq:super}) because  $\lambda'$ is evaluated only when the inner planet transits.  

An important feature of these expressions is the fact that they depend on the two planets' eccentricities only 
through the combination ${\cal Z}$. 
  While that is trivially true for $\delta t_{\cal F}$, 
in  Appendix \ref{sec:zsection} we show that the same is true for $\delta t_{\cal S}$ to a good approximation, due to an apparently
coincidental relationship between different  Laplace coefficients\footnote{
We also show  in Appendix \ref{sec:zsection} that  ${\cal Z}$ is nearly independent of $J$, so that
 if two nearby first order MMR's both contribute comparable  ${\delta t}_{\cal F}$'s,  the eccentricities still
 enter through the single quantity ${\cal Z}$.
}. 
Furthermore, the outer planet's TTV also depends on the same combination ${\cal Z}$.
  As a result, 
  ${\cal Z}$ can 
 be determined quite accurately from TTV's.  Conversely, even if TTV's are well-measured, 
 it is nearly impossible to disentangle the individual planets' eccentricities.
 There is an important exception, however, if the pair of planets is close to the 2:1 resonance (or to the 3:1). In these cases 
 the indirect term leads to a dependence on the individual $z$'s of the two planets.\footnote{
  Kepler-9 \citep{Holman:2010db,borsato2014trades,2014arXiv1403.1372D}, Kepler-18 \citep{Cochran:2011kg}, and Kepler-30 \citep{2012Natur.487..449S} for which relatively precise individual planet eccentricities measurements are reported  contain planets near the 2:1 MMR. 
 }. The presence of additional planets typically does little to  alleviate the degeneracy of inferring individual planet eccentricities.\footnote{ Three (or more) planets with mutual TTV's in principle yield three distinct ${\cal Z}$'s, one for each pairwise interaction, which can be inverted to determine three individual $z$'s. However, the interactions of the most widely separated pair are typically too weak to  constrain their combined ${\cal Z}$. Furthermore, the linear transformation from individual $z$'s to combined ${\cal Z}$'s is nearly singular.}

In addition to the degeneracy between $z$ and $z'$ discussed above, there is a second degeneracy:
between $|{\cal Z}|$ and $\mu'$.   For planet pairs in which only the fundamental TTV is well-measured, 
that degeneracy is evident from Eq. \ref{eq:ttv1X}, since a smaller $\mu'$ can be compensated
by a larger $|{\cal Z}|$ without affecting $\delta t_{\cal F}$.  Moreover, that degeneracy is not in general removed by observing the outer planet's $\delta t_{\cal F}'$, because it depends on the additional unknown $\mu$.
(For a more detailed discussion, as well as a way to break the degeneracy with a statistical sample of TTV's  see LXW and \citet{2014ApJ...787...80H}).
However, the $|{\cal Z}|$-$\mu$ degeneracy {\it can} be broken if both fundamental and chopping TTV's are observed \citep{Deck:2014vr}, or if  fundamental and secondary TTV's are observed.  
We give examples below.

\section{Four Systems}
\label{sec:systems}

We analyze the TTV's in four planetary systems  observed by the {\it Kepler} telescope, compriseing ten planets.  
Our analysis is based on the transit times computed by \citet{2015ApJS..217...16R}, which incorporates observations  from Quarters 1-17.
These four systems  exhibit clear TTV's that  have not yet been analyzed in detail.

\subsection{Methods}
\label{sec:methods}

We employ three complementary methods:
\begin{itemize}
	\item {\bf N-body MCMC}: Our setup is fairly standard, and is described in Appendix \ref{sec:nbody_mcmc}.
	Our default priors are
	  logarithmic in masses ($dP/dM \propto 1/M$) and uniform  in eccentricity ($dP/de \propto $ const).  Planet densities inferred from TTV's are often surprisingly low (see below and  \citet{Wu:2013cp,2014ApJ...787...80H,Weiss:2014ef}).  Therefore 
	   we also employ a
	 second set of ``high mass priors''  that are
        uniform in masses ($dP/dM\propto$ const.) and logarithmic in eccentricity ($dP/de\propto 1/e$), 
        where  the latter weights more towards lower eccentricity, 
        and consequently also towards high masses via Equations (\ref{eq:ttv1X}) and (\ref{eq:ttv3X}).
   	
	\item {\bf Analytic MCMC}: We  run MCMC simulations that model the TTV with  analytic 
	formulae (Eqs. \ref{eq:ttv1X}-\ref{eq:ttv3X}), rather than with N-body simulations.  Details
	are provided in Appendix \ref{sec:analytic_mcmc}.  The analytic MCMC results agree well
	with the N-body ones for the systems considered in this paper (see below). This 
	provides support for the analytic model and, more importantly, shows that the inferred
	 planet parameters can be understood with the help of the analytic model.

	\item {\bf Analytic Constraint Plot}:  We use the analytic formulae to infer how
	 each of the 
	TTV components (fundamental, chopping, and secondary) constrains the masses and eccentricities, 
	and thereby show how the overlapping  constraints explain the MCMC results. 
	To do so, we first fit for the amplitudes of the sinusoids in Eqs. \ref{eq:ttv1X}-\ref{eq:ttv3X}.
	For the inner planet's TTV, there are five unknowns to be fit for:  (a)  the complex amplitude of  the sinusoid with period equal to the superperiod (Eq. \ref{eq:ttv1X}), or equivalently
	the real amplitudes of the sine and cosine component; (b) the real amplitude of
	the infinite sum of sinusoids in Eq. \ref{eq:ttv2X} (noting that the  phase of this term is known)	; 
	and (c) the complex amplitude of the secondary TTV.  Since the time dependence
	of each of these terms is known, the fit is done with a simple linear least squares. 

	Next, setting the complex amplitude inferred from (a) equal to $\mu'(A+
	B{\cal Z})$, we solve for $|{\cal Z}|$ as a function of $\mu'$. The result is
	a line in the $\mu'-|{\cal Z}|$ plane that is allowed by the fundamental TTV.  Accounting
	for the observational errors turns the line into a band.  
	Similarly, the  amplitude of $\delta t_{\cal C}$ constrains $\mu'$, and the complex
	amplitude of $\delta t_{\cal S}$ provides another band of constraint in the $\mu'-|{\cal Z}|$
	plane.

\end{itemize}

\begin{figure}
\begin{center}
\includegraphics[width=0.45\textwidth]{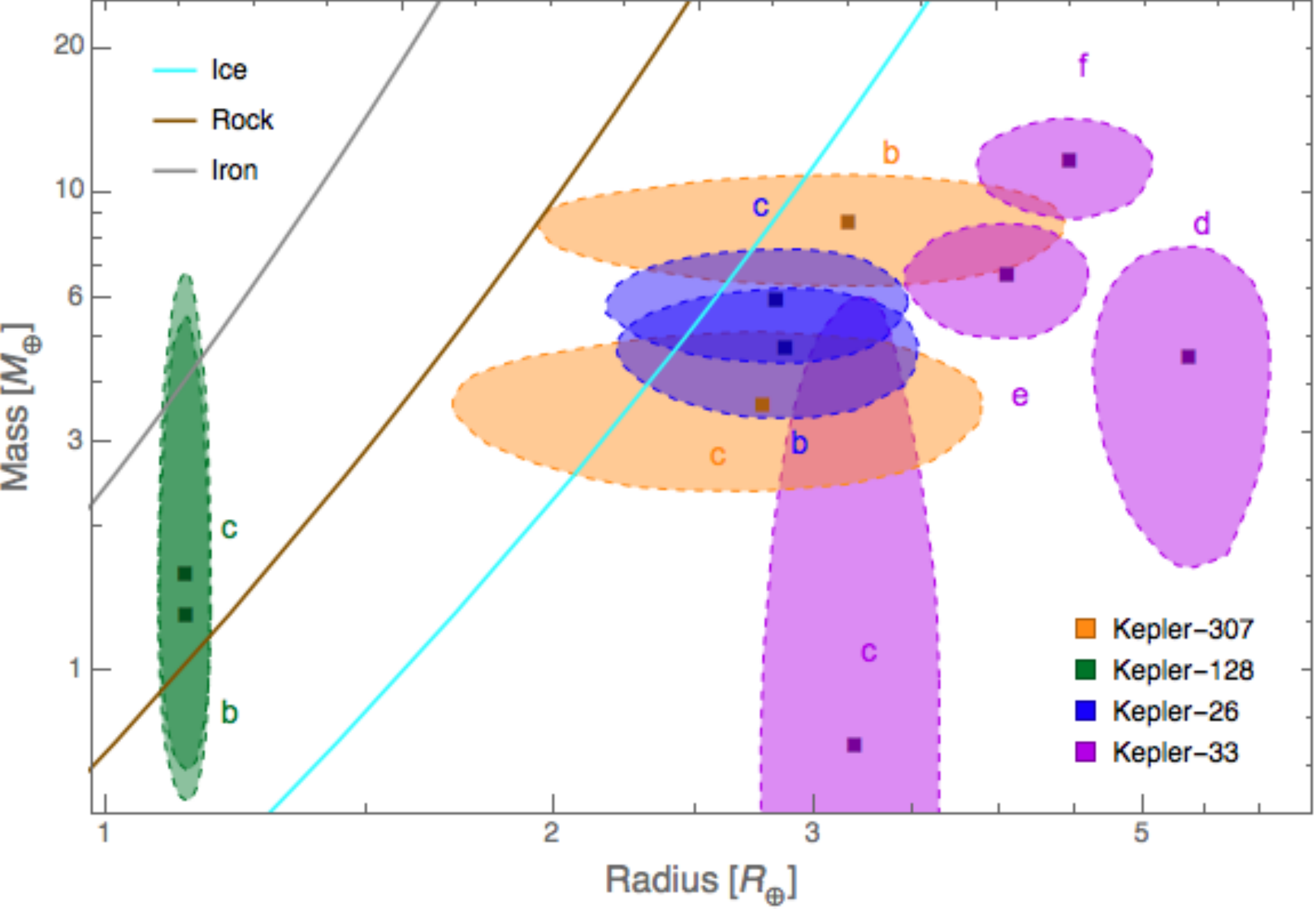}
\includegraphics[width=0.45\textwidth]{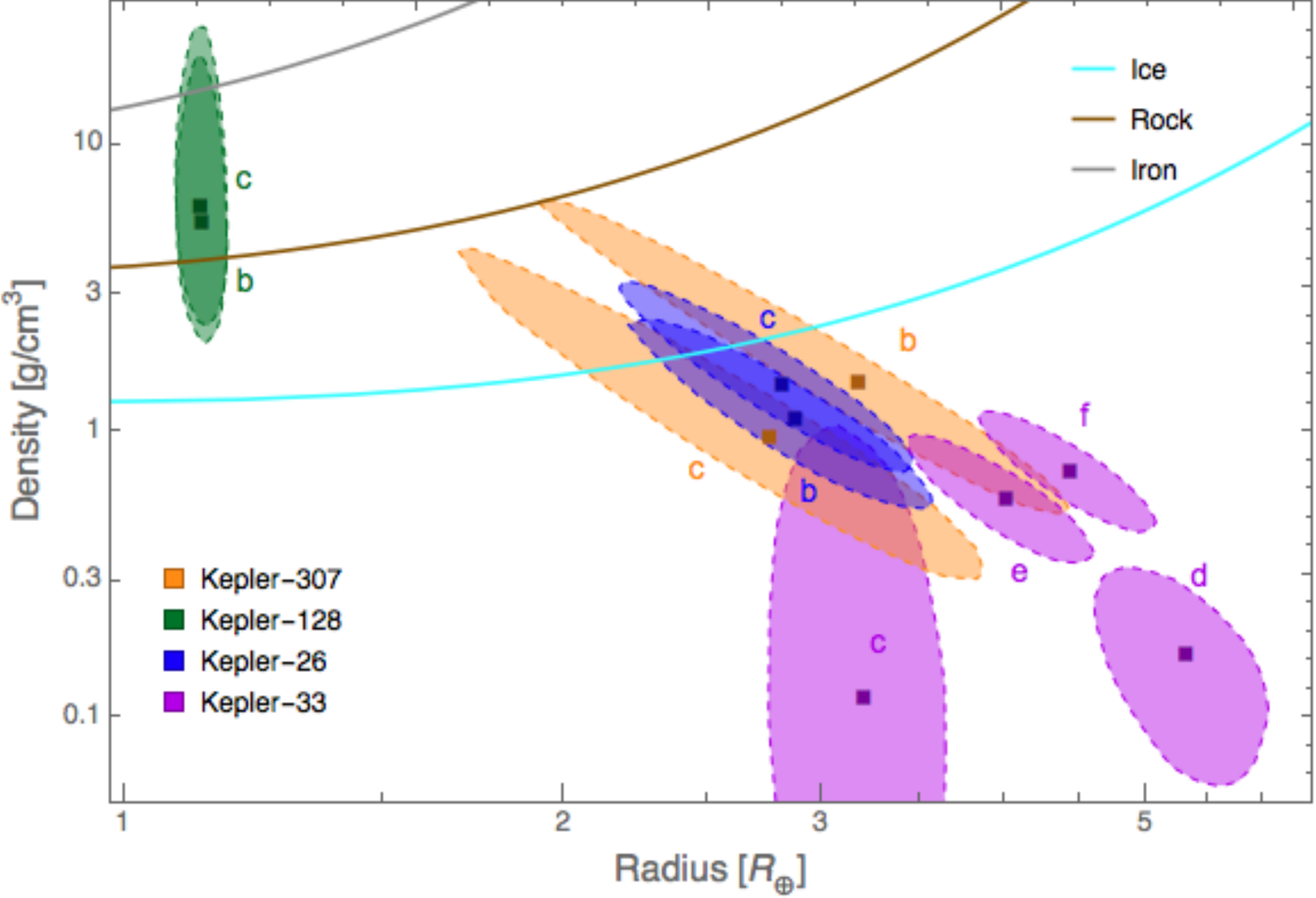}
\caption{ 
{\bf Top panel: } Planet mass versus radius for each planet presented in Section \ref{sec:systems}. 
Each  splotch shows the  68\% joint confidence region in mass and radius. 
Theoretical mass-radius relationships for planets composed of pure ice, rock, and iron from \citet{Fortney:2007hf} are plotted as colored curves. 
Confidence regions are 
from the MCMC results, combined with the Exoplanet Archive values of stellar mass and planet radius (accounting for their errors by drawing samples from Gaussian distributions).
{\bf Bottom panel: } Same as top panel except with density plotted on the vertical axis.
}
\label{fig:MassRadius}
\end{center}
\end{figure}

In the following  subsections, we describe our results for each of the  four systems.
All of the inferred planet masses and densities are summarized in  Table \ref{tab:mass_radius}
and Figure \ref{fig:MassRadius}. 
In Table \ref{tab:mass_radius}, and throughout this paper, 
measured values refer to the median.  
The upper and lower error bars demarcate the zone of 68\% confidence (`1-sigma') that is bounded by the 84\% and 16\% quantiles, respectively.

The eccentricity results are 
 in Table \ref{tab:ecc}.
We focus on inferring the combined eccentricity
$|{\cal Z}|\approx |z'-z|/\sqrt{2}$ rather than  $z$ and $z'$ individually, 
which
 are nearly impossible to disentangle from one another.
 We expect that  $|{\cal Z}|$ is typically a good surrogate for the individual planets' 
 eccentricities.  
  However, it is conceivable that $z\approx z'$, i.e.,  the two planets have comparable eccentricities and aligned orbits. If so, the individual eccentricities could be much higher
than $|{\cal Z}|$.  Such a situation could arise if damping has acted on the planetary system, removing one of the secular modes but not the other. Although we do not favor that scenario, it remains a possibility that is difficult to exclude.

\begin{table*}
\caption{Planet Properties}
\begin{center}
\begin{tabular}{|c c c c  | c c|}
\hline
{\bf Planet}	& {\bf Period}	& {\bf Radius}	&{\bf Stellar Mass}	& {\bf Mass}	& \bf{ Density}\\ 
	&		 [days]	& $[R_\oplus]$		& $[M_{\odot}]$		& $[M_\oplus]$	& [g/$\text{cm}^3$]\\ \hline
Kepler-307b	& 10.42	& $3.2^{+1.2}_{-0.5}$&	$0.98^{+0.14}_{-0.09}$	& $8.6^{+1.6}_{-1.4}$	& $1.5^{+1.7}_{-0.7}$\\ 
Kepler-307c	& 13.08	& $2.8^{+1.0}_{-0.4}$&	---					& $3.7^{+1.0}_{-0.8}$	& $0.9^{+1.1}_{-0.5}$\\ \hline
Kepler-128b	& 15.09	& $1.13^{+0.03}_{-0.03}$&$1.18^{+0.07}_{-0.07}$	& $1.3^{+1.9}_{-0.6}$	& $5.0^{+7.1}_{-2.3}$\\ 
Kepler-128c	& 22.80	& $1.13^{+0.03}_{-0.03}$&--- 	& $1.5^{+2.2}_{-0.7}$	& $5.8^{+8.5}_{-2.8}$\\ \hline
Kepler-26b	& 12.28	& $2.9^{+0.4}_{-0.4}$&	$0.55^{+0.08}_{-0.08}$		& $4.8^{+0.8}_{-0.8}$	& $1.1^{+0.5}_{-0.5}$\\ 
Kepler-26c	& 17.25	& $2.8^{+0.4}_{-0.4}$&	--- 		& $6.0^{+0.8}_{-0.8}$	& $1.5^{+0.7}_{-0.7}$\\ \hline
Kepler-33c	& 13.18	& $3.2^{+0.3}_{-0.3}$&	$1.29^{+0.06}_{-0.12}$			& $0.8^{+2.5}_{-0.7}$	& $0.1^{+0.4}_{-0.1}$\\ 
Kepler-33d	& 21.78	& $5.4^{+0.5}_{-0.5}$&	---			& $4.7^{+2.0}_{-2.0}$	& $0.2^{+0.1}_{-0.1}$\\ 
Kepler-33e	& 31.78	& $4.0^{+0.4}_{-0.4}$&	---			& $6.7^{+1.2}_{-1.3}$	& $0.6^{+0.2}_{-0.2}$\\ 
Kepler-33f		& 41.03	& $4.5^{+0.4}_{-0.4}$&	---			& $11.5^{+1.8}_{-2.1}$	& $0.7^{+0.2}_{-0.2}$ \\ \hline
\end{tabular}
\end{center}
\tablecomments{Properties of each TTV planet considered in this paper. The period, radius, and stellar mass are from  the Exoplanet Archive\footnote{\url{http://exoplanetarchive.ipac.caltech.edu}} \citep{akeson2013archive}, and  the mass and density result  from N-body MCMC simulations.
The quoted planet mass and density incorporate the stellar mass and planet radius, with errors added in quadrature. 
}
\label{tab:mass_radius}
\end{table*}%

\begin{table}
\tabletypesize{\scriptsize}
\caption{Eccentricities}
\begin{center}
\begin{tabular}{|c c c c|}
\hline
{\bf Planet Pair}&{\bf Resonance}& $\Delta$	&  $|{\cal Z}|$ \\  \hline
Kepler-307b/c	& 5:4	& 0.0050	& $0.0017^{+0.0005}_{-0.0004}$\\ 
Kepler-128b/c	& 3:2	& 0.0075	& $0.06^{+0.04}_{-0.03}$ \\
Kepler-26b/c	& 7:5	& 0.0032	& $0.013^{+0.002}_{-0.005}$\\ 
Kepler-33c/d	& 5:3	& -0.0084	& $0.03^{+0.02}_{-0.02}$\\ 
Kepler-33d/e	& 3:2	& -0.0269	& $0.010^{+0.005}_{-0.004}$\\ 
Kepler-33e/f	& 9:7	& 0.0040	& $0.006^{+0.002}_{-0.002}$\\  \hline
\end{tabular}
\end{center}
\tablecomments{The median combined eccentricities from N-body MCMC,  along with 1-$\sigma$ uncertainties. The nearest first- or second-order MMR to the pair's period ratio is given along with the pair's distance to resonance, $\Delta$ (i.e., Eq. \ref{eq:delta_defn} for the first order resonances, and its obvious extension for the
second order ones: $\Delta\equiv{(K-2) \over K}{P'\over P}-1$). }
\label{tab:ecc}
\end{table}

\subsection{Kepler-307 (KOI-1576)}
\label{sec:kep307}

\begin{figure}[htbp]
	\begin{center}
	\includegraphics[width=0.45\textwidth]{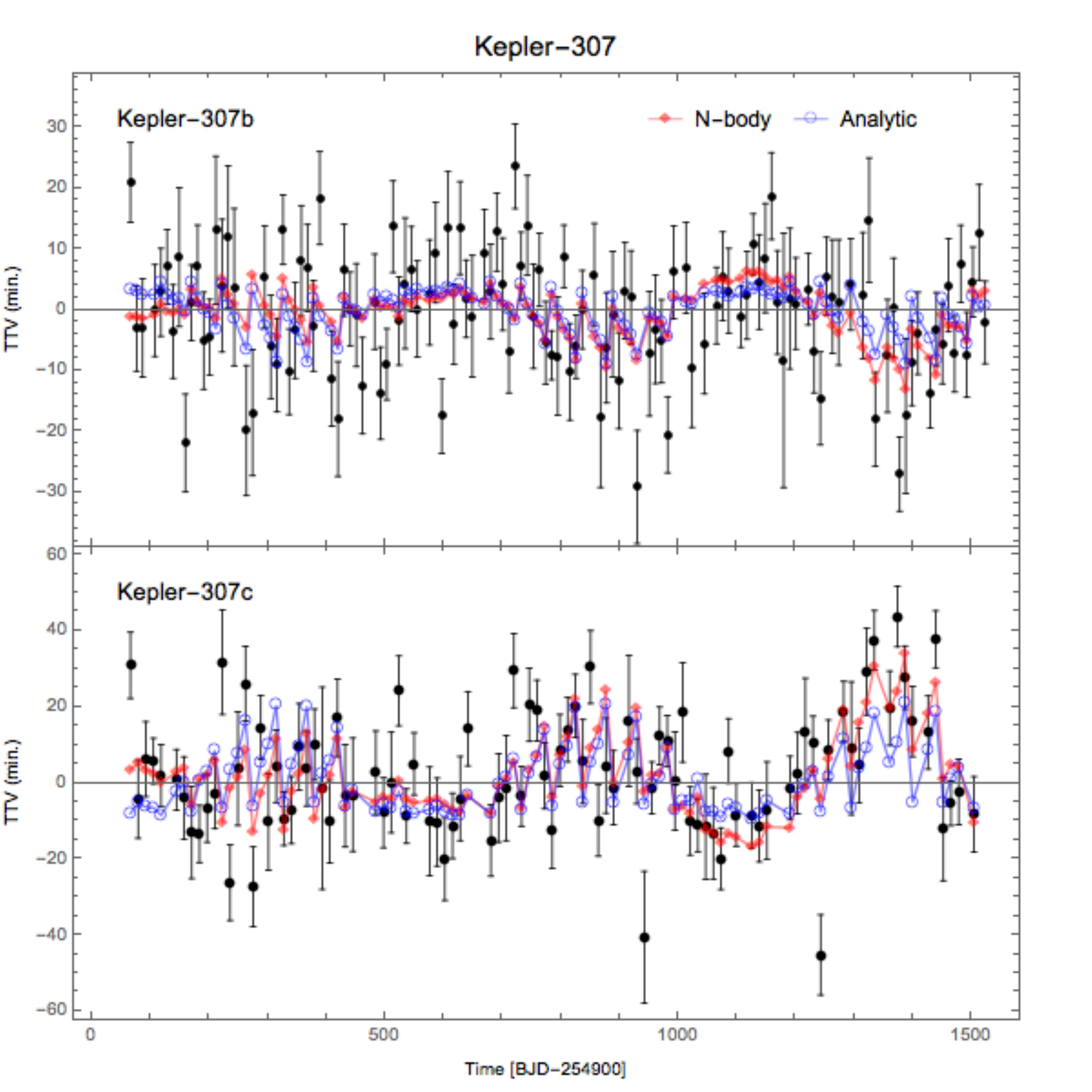}
\caption{
	The TTVs of Kepler-307b and c and their uncertainties, computed by \citet{2015ApJS..217...16R} are shown as black points with error bars. The best-fit N-body solution from MCMC fitting is plotted as red diamonds.  The best-fit analytic model solution is plotted as blue circles.
}
	\label{fig:kep307ttv}
	\end{center}
\end{figure}

Kepler-307b and c are a pair of sub-Neptune sized planets, with radii of $R_b=3.2^{+1.2}_{-0.5}~R_\oplus$ and $R_c=2.8^{+1.1}_{-0.4}~R_\oplus$. 
The pair were confirmed as planets by \citet{xie2014transit} on the basis of their TTVs.
 The pair's orbits are near a 5:4 MMR with $\Delta=0.005$.  The planets' TTVs are shown in Figure \ref{fig:kep307ttv} along with the best-fit N-body and analytic solutions for the transit times. 
One can see both the low  frequency fundamental TTV, as well as the high frequencies from the chopping TTV.

\begin{figure}[htbp]
	\begin{center}
	\includegraphics[width=0.45\textwidth]{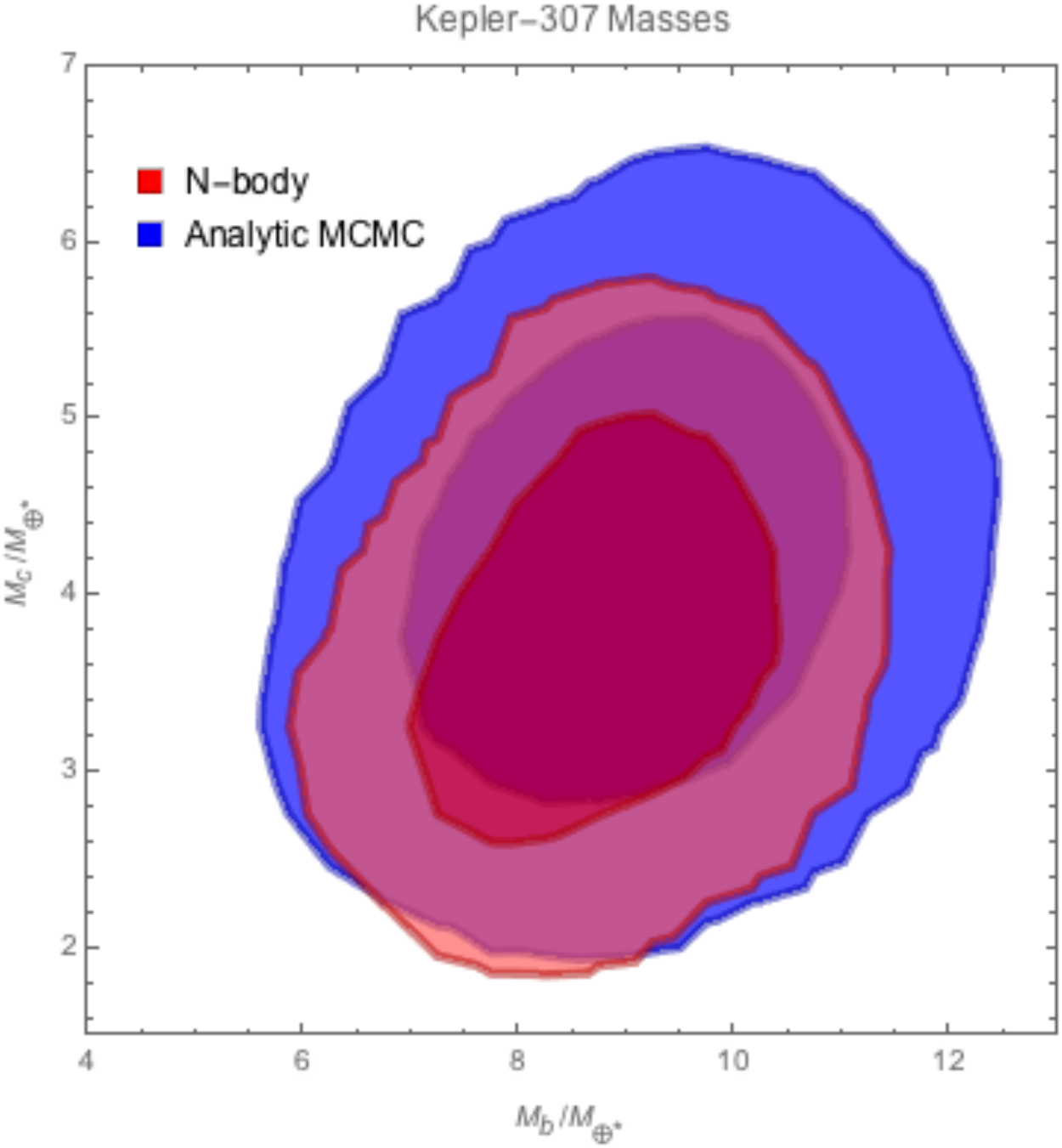}
\caption{
N-body (red) and analytic (blue) MCMC posterior distributions in planet mass for the Kepler-307 system.  
The dark  and light shading indicate the 68\% and 95\% confidence regions, respectively.
}
\label{fig:kep307mass}
\end{center}
\end{figure}

\begin{figure}[htbp]
	\begin{center}
	\includegraphics[width=0.45\textwidth]{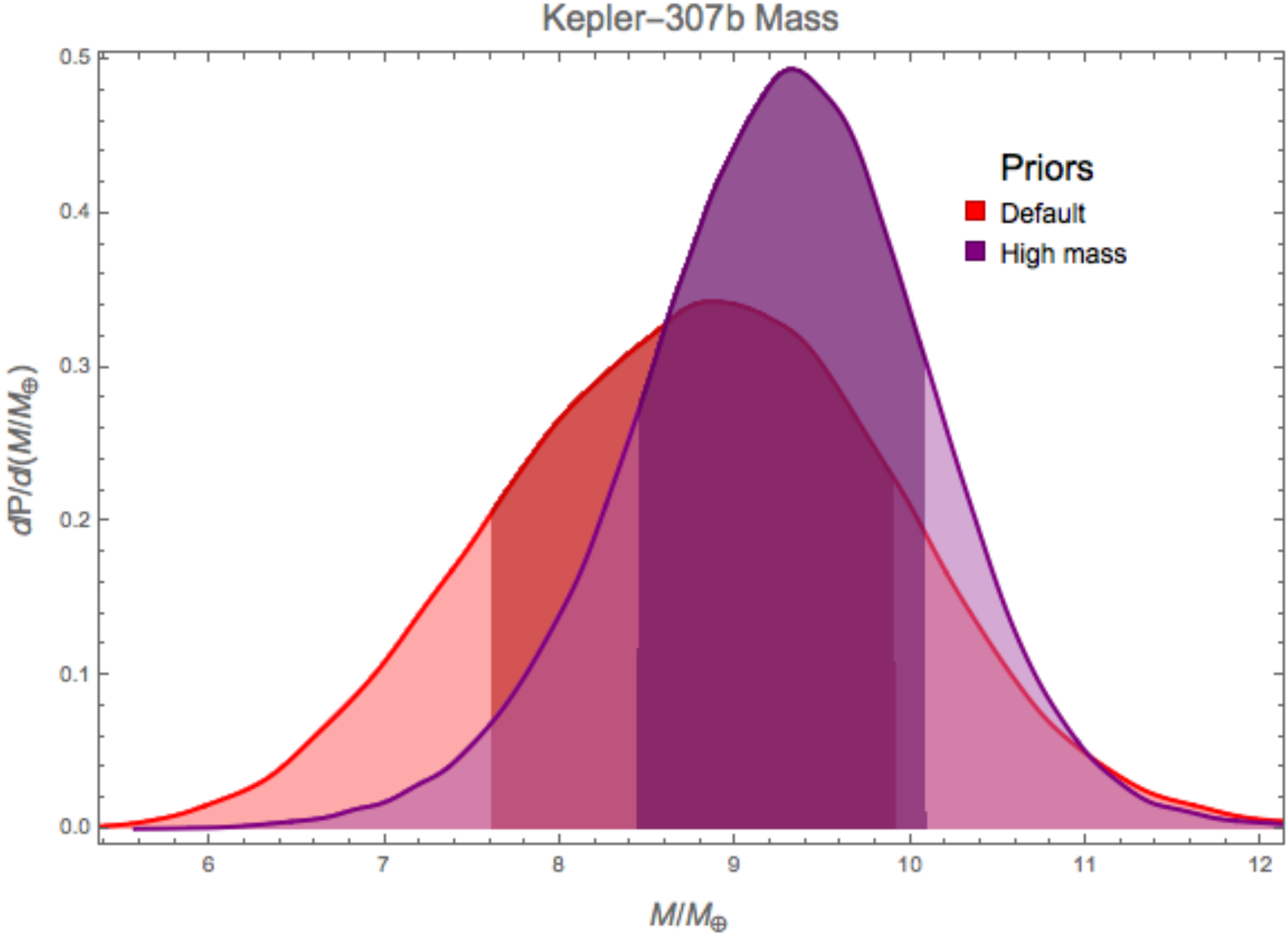}
	\includegraphics[width=0.45\textwidth]{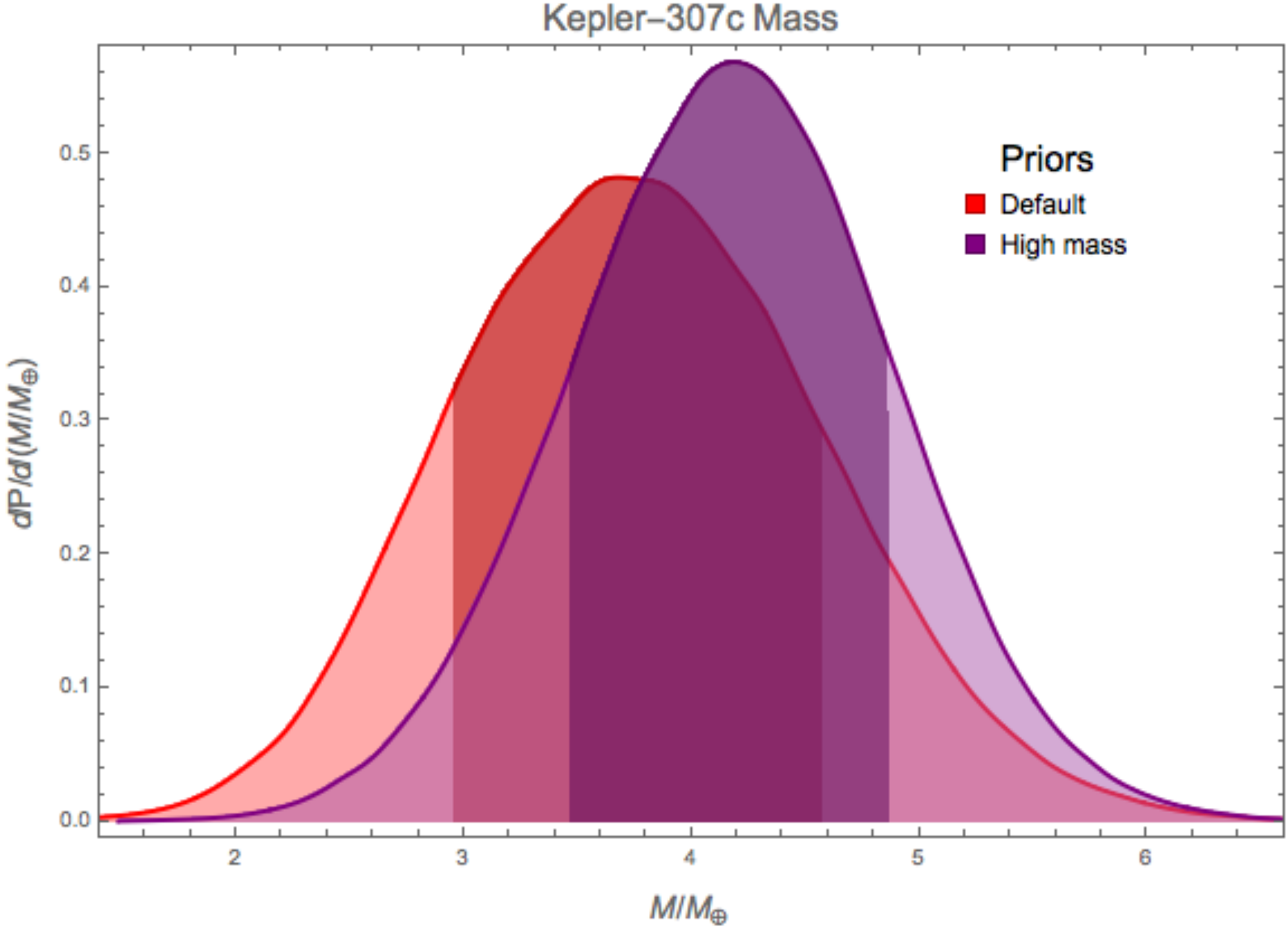}
\caption{
Posterior distribution of planet-to-star mass ratios from N-body MCMC using the default and `high mass' priors (see Section \ref{sec:methods}).
Each curve is a Gaussian kernel density estimate of the corresponding posterior sample.
The  68\% equal-tailed credible regions in mass are emphasized by darker shading. 
}
	\label{fig:kep307mass_prior}
	\end{center}
\end{figure}

For the N-body MCMC, an ensemble of 800 walkers was evolved for 250,000 iterations, saving every 800th iteration. A resulting $\sim 26,000$ independent posterior samples were generated based on analysis of the walker auto-correlation lengths (Appendix \ref{sec:nbody_mcmc}). The joint posterior distribution of planet masses from analytic and N-body MCMC are shown in Figure \ref{fig:kep307mass}.
The methods show excellent agreement.
Note that the MCMC constrains $\mu$ (the ratio of planet to star mass), and so
 masses in the figure are in units of
\begin{eqnarray}
\label{eq:m_earth_star}
M_{\oplus *}\equiv M_\oplus\times\frac{M_*}{M_\odot} \ ,
\end{eqnarray}
which differs slightly from an Earth mass. 
Figure \ref{fig:kep307mass_prior} compares N-body MCMC posteriors computed using our default priors and and high mass priors (see Section \ref{sec:methods}). 
The inferred planet masses are not strongly effected by the choice of priors.

	
\begin{figure}
	\begin{center}
	\includegraphics[width=0.45\textwidth]{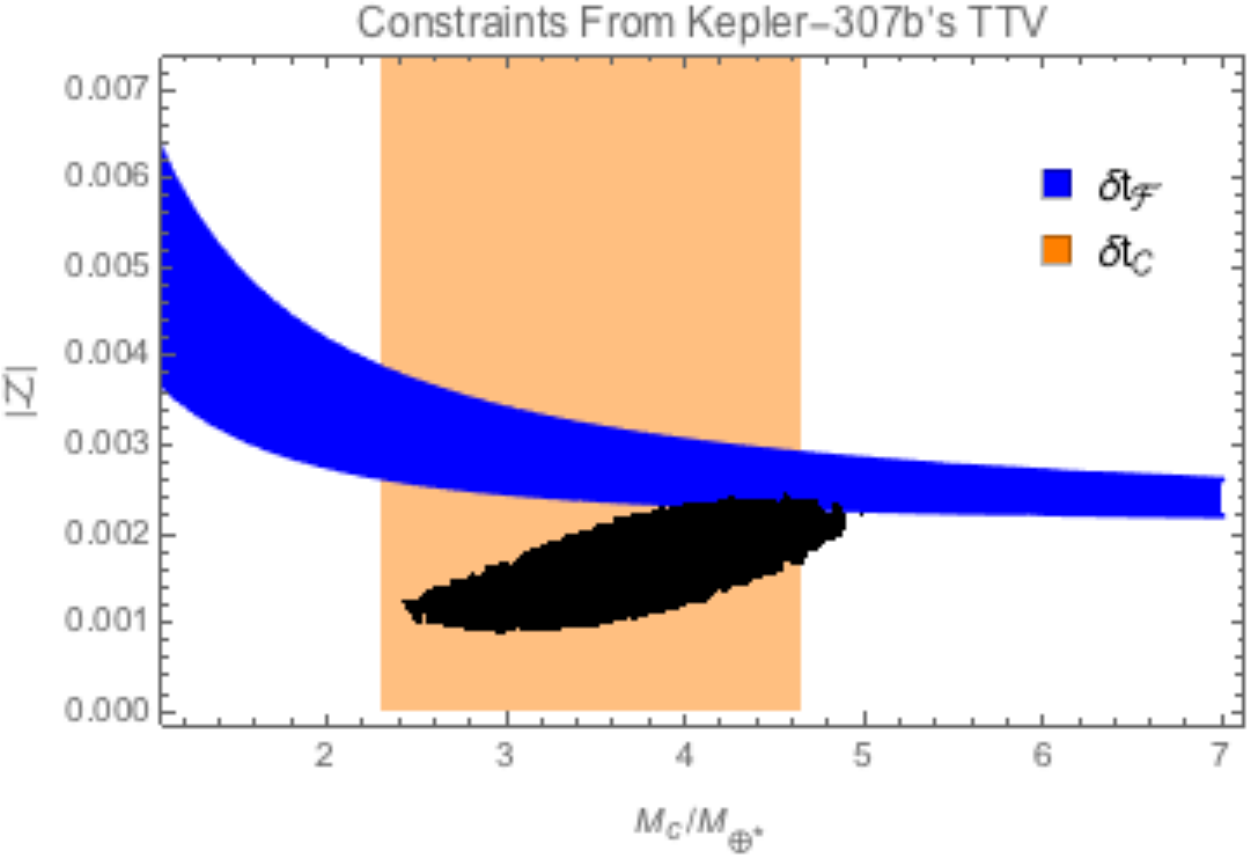}
    \includegraphics[width=0.45\textwidth]{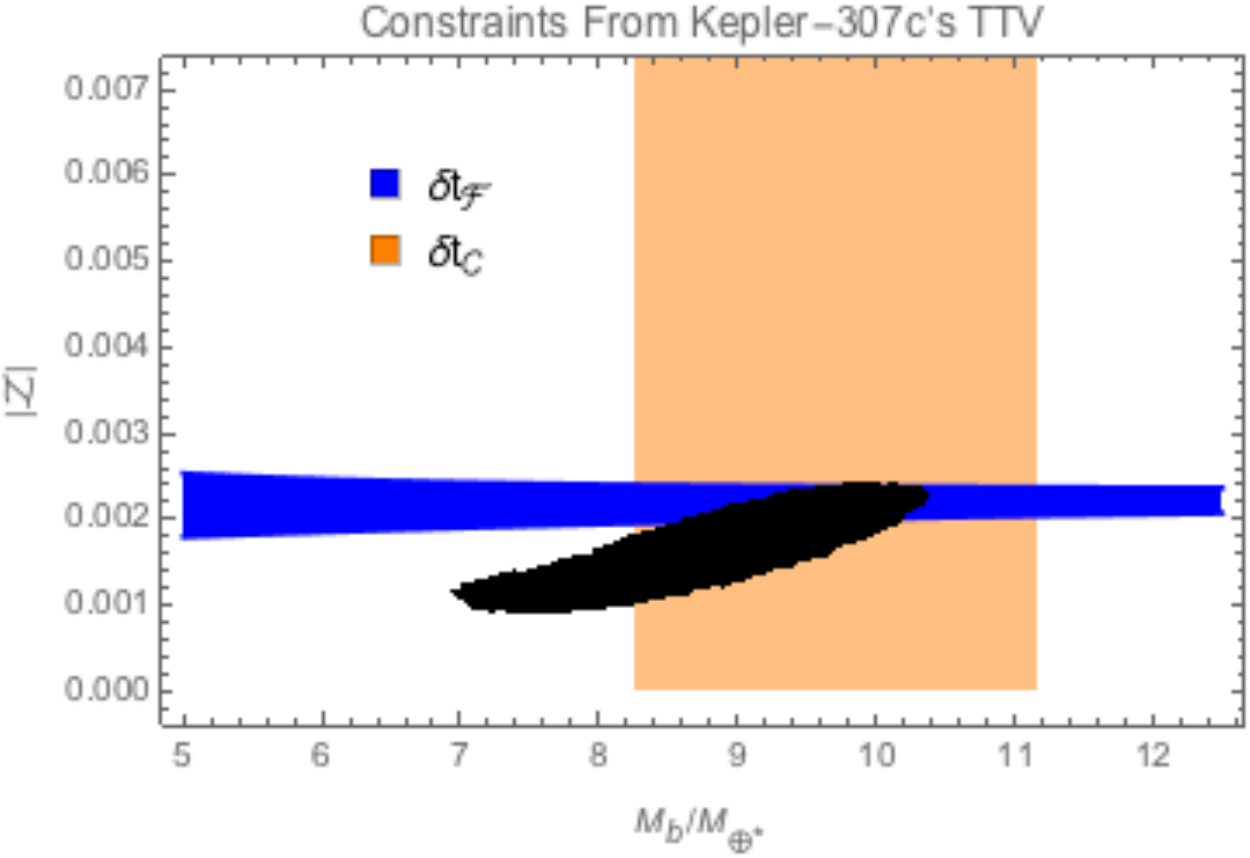}
	\caption{
	Analytic Constraint Plots for Kepler-307: The top panel shows the constraints provided by 
	the  inner planet's fundamental and chopping TTV amplitudes  (see Section \ref{sec:analytic_model}). 
	The black regions shows the N-body MCMC result, at 68\% confidence. The bottom panel 
	shows the same for the outer planet.  
}
	\label{fig:kep307constraints}
	\end{center}
\end{figure}

Figure \ref{fig:kep307constraints} shows  the analytic  constraint plots  (Section \ref{sec:methods})
for the inner and outer planets. 
The MCMC result is roughly consistent with where the constraints from the fundamental
and chopping components overlap. 
	Hence  those two components
	 are  primarily responsible for  this system's inferred masses and eccentricities.

\begin{figure}[htbp]
	\begin{center}
	\includegraphics[width=0.45\textwidth]{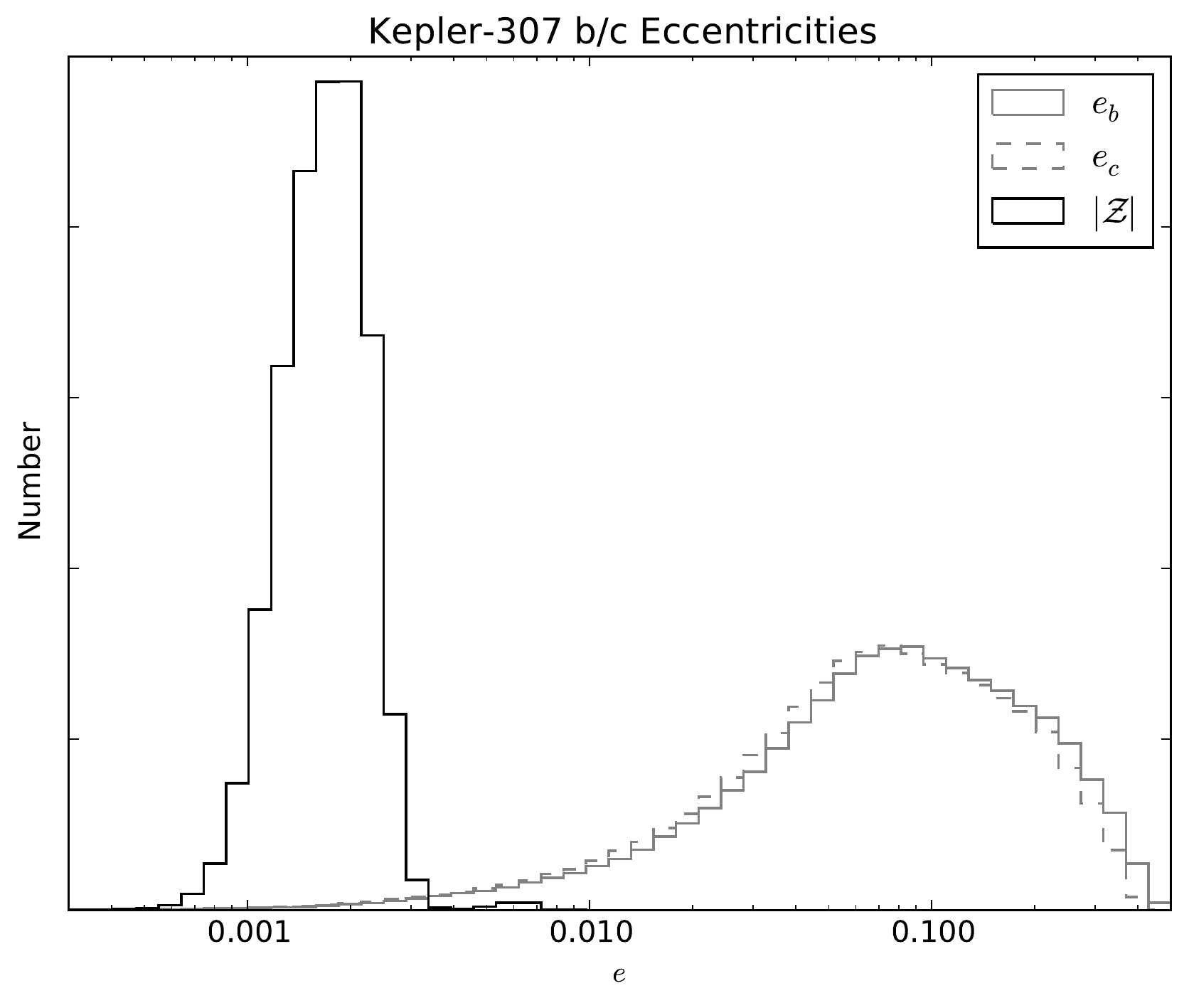}
	\caption{
 Histograms of the eccentricities of Kepler 307b and c, along with the combined eccentricity $|{\cal Z}|$, as computed by N-body MCMC. 
 The individual eccentricities of planet `b' and `c' are poorly constrained from the TTV while ${\cal Z}$ is measured accurately, as predicted by the analytic model.
}
	\label{fig:kep307ecc}
	\end{center}
\end{figure}

Figure \ref{fig:kep307ecc} illustrates that the combined eccentricity variable $|{\cal Z}|$ is  inferred much more accurately  than the individual planets' eccentricities (Section \ref{sec:inferring}). The plot shows
 the posterior distributions of the individual planet eccentricities from the N-body MCMC, as
well as that of $|{\cal Z}|$.  The eccentricities of planets `b' and `c' are essentially unconstrained by the TTVs and show a nearly uniform  distribution for $e\lesssim 0.1$. (Note that the x-axis axis in the figure is logarithmic.) By contrast, the  distribution of  $|{\cal Z}|$ is sharply peaked around $|{\cal Z}| \sim 0.002$. The situation illustrated by Figure \ref{fig:kep307ecc} is typical of the MCMC results for all systems in this paper: only $|{\cal Z}|$  is well-constrained, not the invidividual planets' eccentricities, which largely reflect the priors.

	The lightcurve of Kepler-307 shows a third  (candidate) planet, KOI 1576.03, with a period of 23.34 days and radius of $\sim 1.2R_\oplus$ that we have ignored in our TTV modeling. The period of this candidate planet places it far from any low order MMRs with the other two planets and its influence on the TTVs of Kepler-307b and c should be negligible, especially given its small size.
\subsection{Kepler-128 (KOI-274)}
\label{sec:kep0274}
\begin{figure}[htbp]
	\begin{center}
	\includegraphics[width=0.45\textwidth]{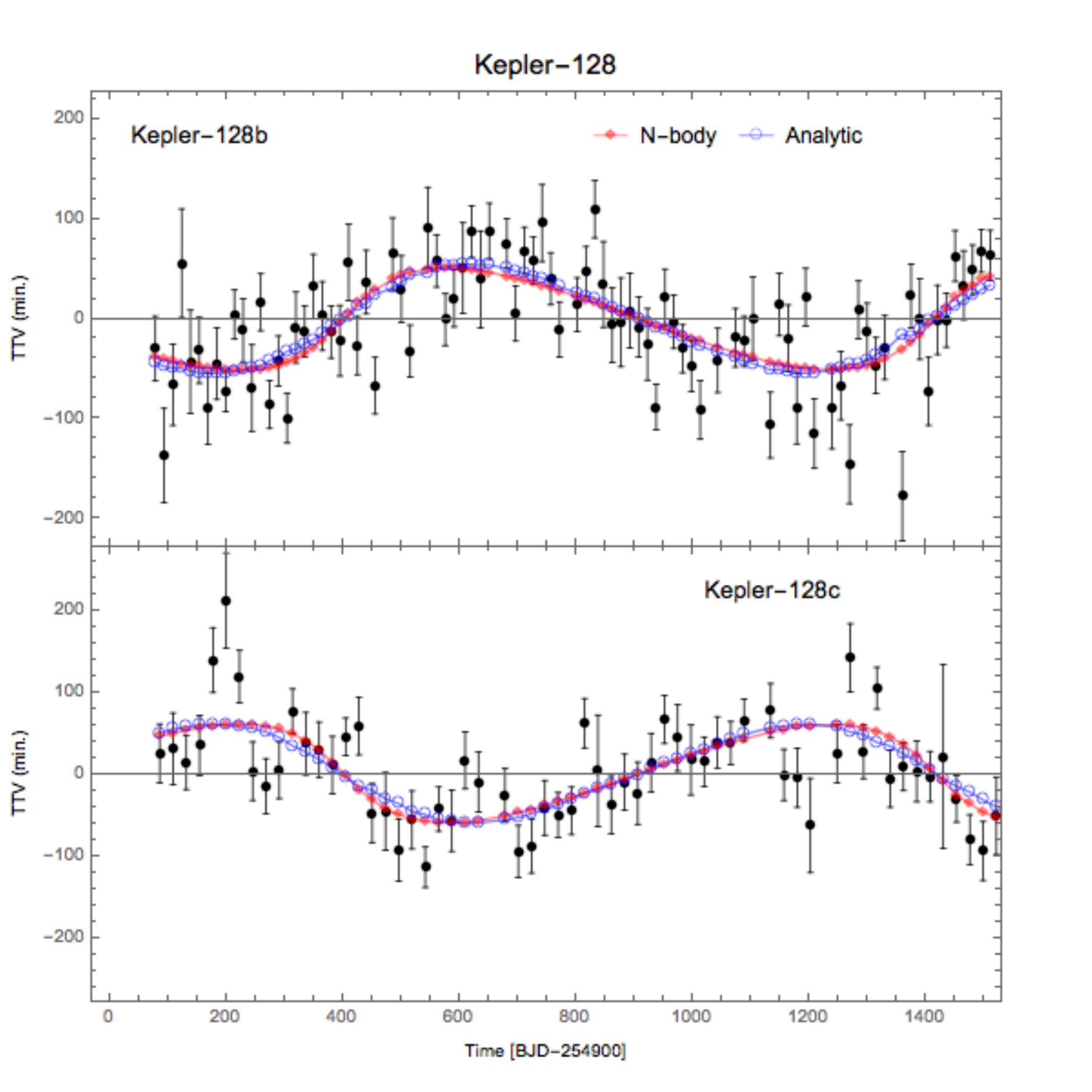}
	\caption{
	TTVs of the Kepler-128 system (see Figure \ref{fig:kep307ttv} for description).
	}
	\label{fig:kep128ttv}
	\end{center}
\end{figure}

Kepler-128b and c are pair of approximately Earth-sized planets  with orbit that place them just wide of the 3:2 MMR ($\Delta = 0.0075$).
The pair were confirmed as planets by \citet{xie2014transit} on the basis of their TTVs. 
The TTVs of Kepler-128b and c are shown in Figure \ref{fig:kep128ttv}.
A non-zero secondary component is present in addition to the fundamental TTV and causes the slight `skewness' in the otherwise sinusoidal TTVs.  

\begin{figure}[htbp]
	\begin{center}
	\includegraphics[width=0.47\textwidth]{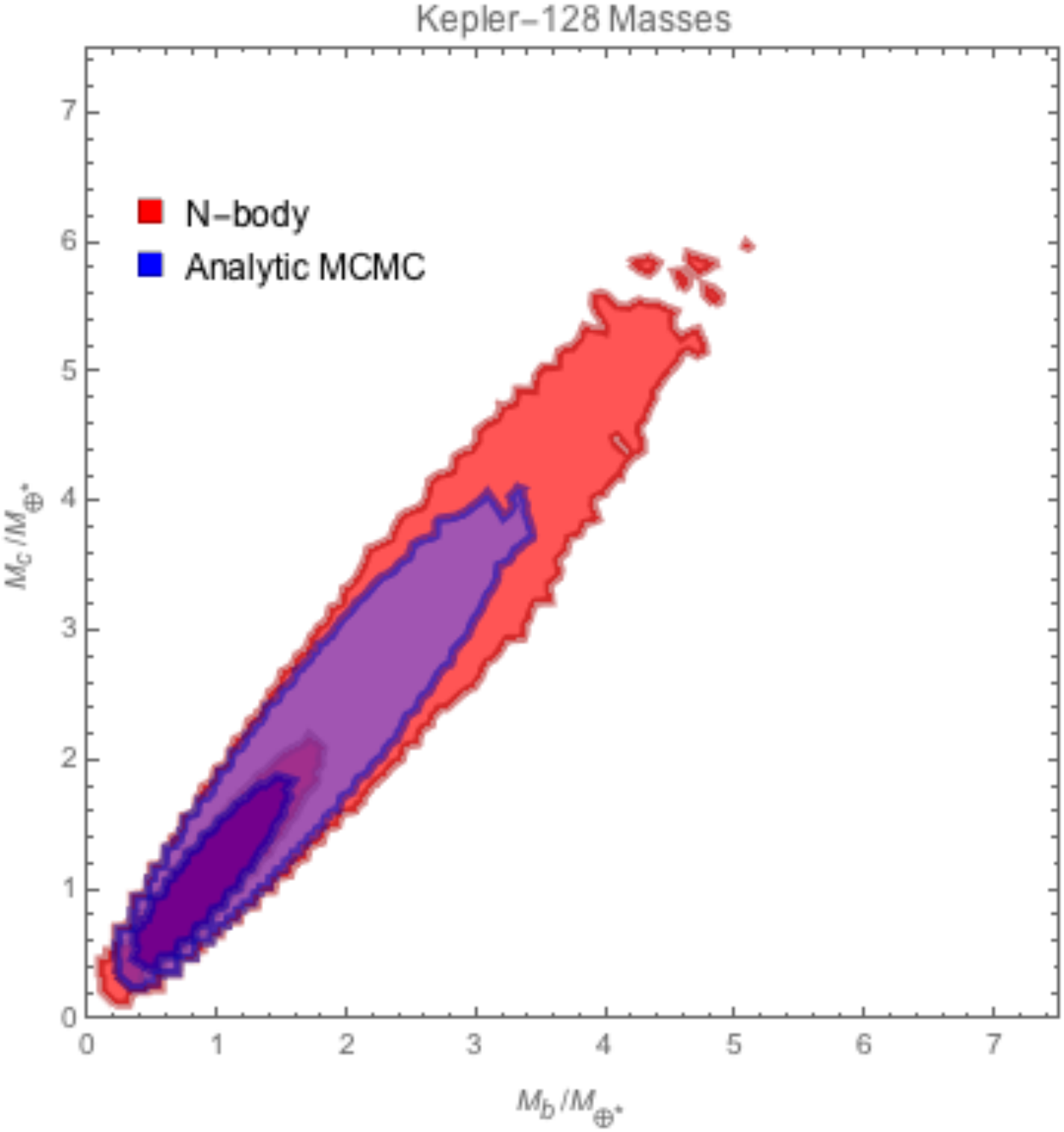}
	\caption{
MCMC mass posterior for the Kepler-128 system (see Figure \ref{fig:kep307mass} for description).
	}
	\label{fig:kep128mass}
	\end{center}
\end{figure}

\begin{figure}[htbp]
	\begin{center}
	\includegraphics[width=0.45\textwidth]{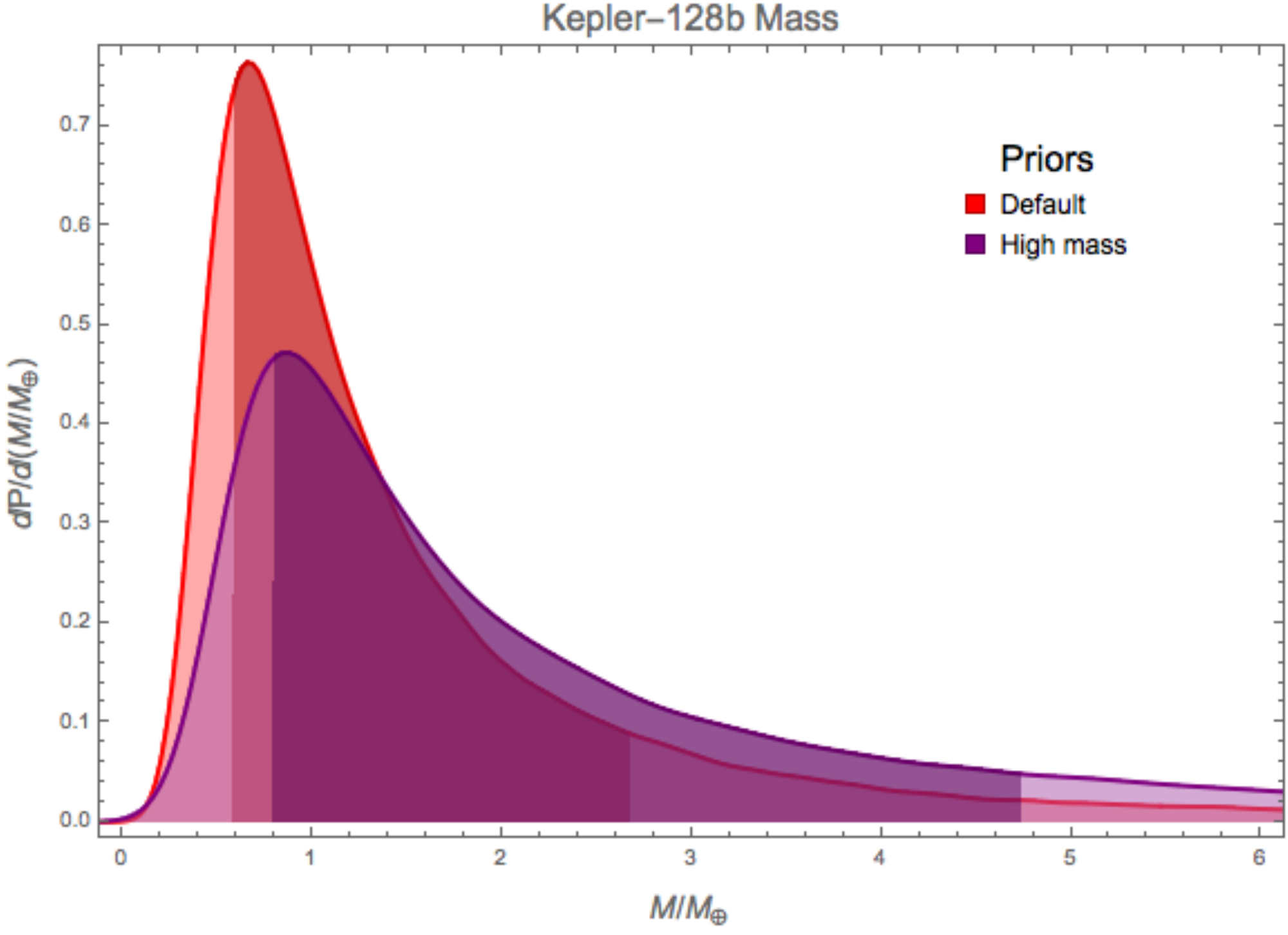}
	\includegraphics[width=0.45\textwidth]{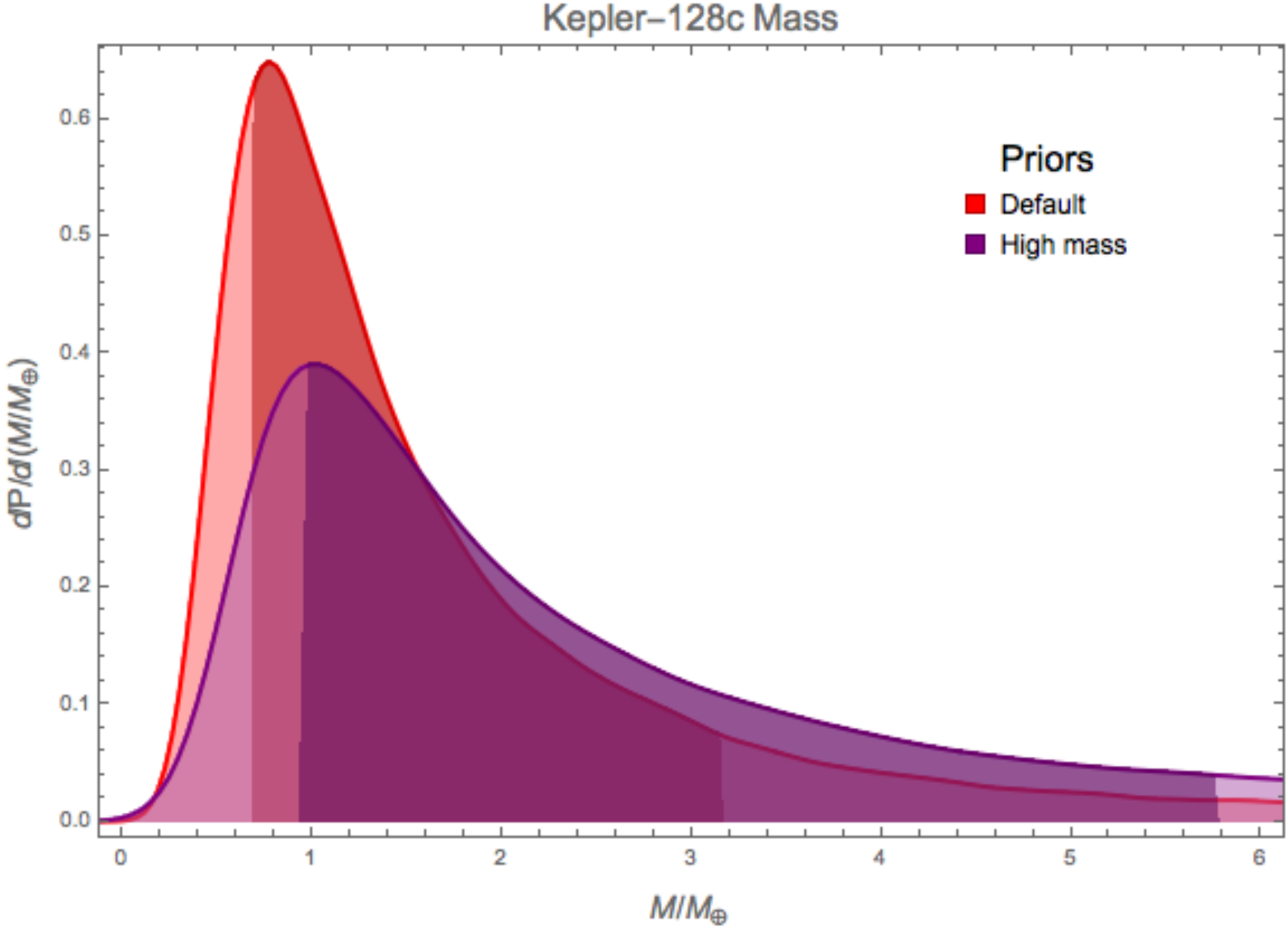}
\caption{
Comparison of MCMC priors for the Kepler-128 system (see Figure \ref{fig:kep307mass_prior} for description).
}
	\label{fig:kep128mass_prior}
	\end{center}
\end{figure}

For the N-body MCMC, an ensemble of 800 walkers was run for 250,000 iterations, saving every 800th iteration, resulting in  $\sim 7,200$ independent posterior samples, based on analysis of the Markov chains' auto-correlation lengths. 
The planet mass constraints derived from MCMC are shown in Figure \ref{fig:kep128mass}.
Figure \ref{fig:kep128mass_prior} compares MCMC results using the default and high mass priors. 
The peaks of the marginal mass posterior distributions remain roughly the same for both priors but more of the posterior probability is shifted to higher mass for the latter choice.


\begin{figure}[htbp]
\begin{center}
	\includegraphics[width=0.45\textwidth]{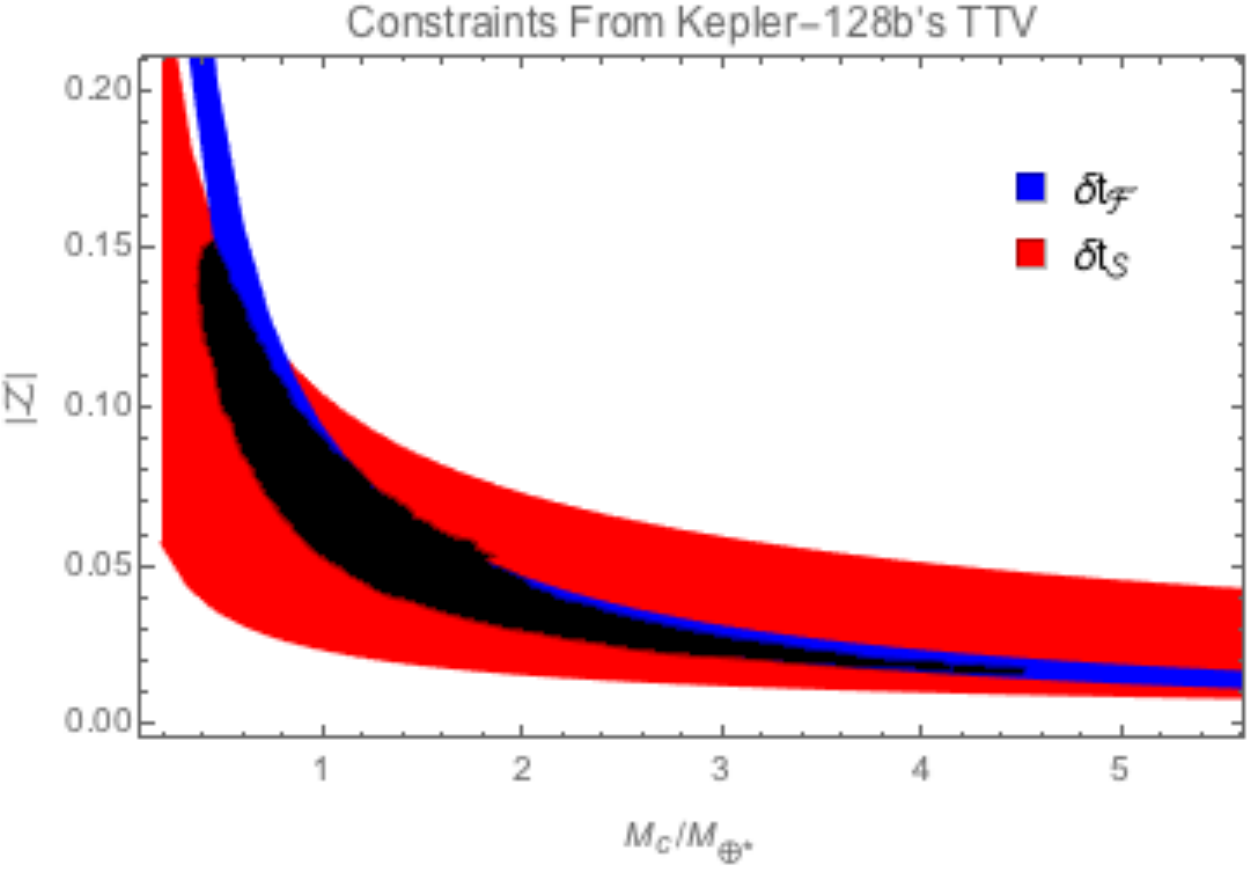}
   	\includegraphics[width=0.45\textwidth]{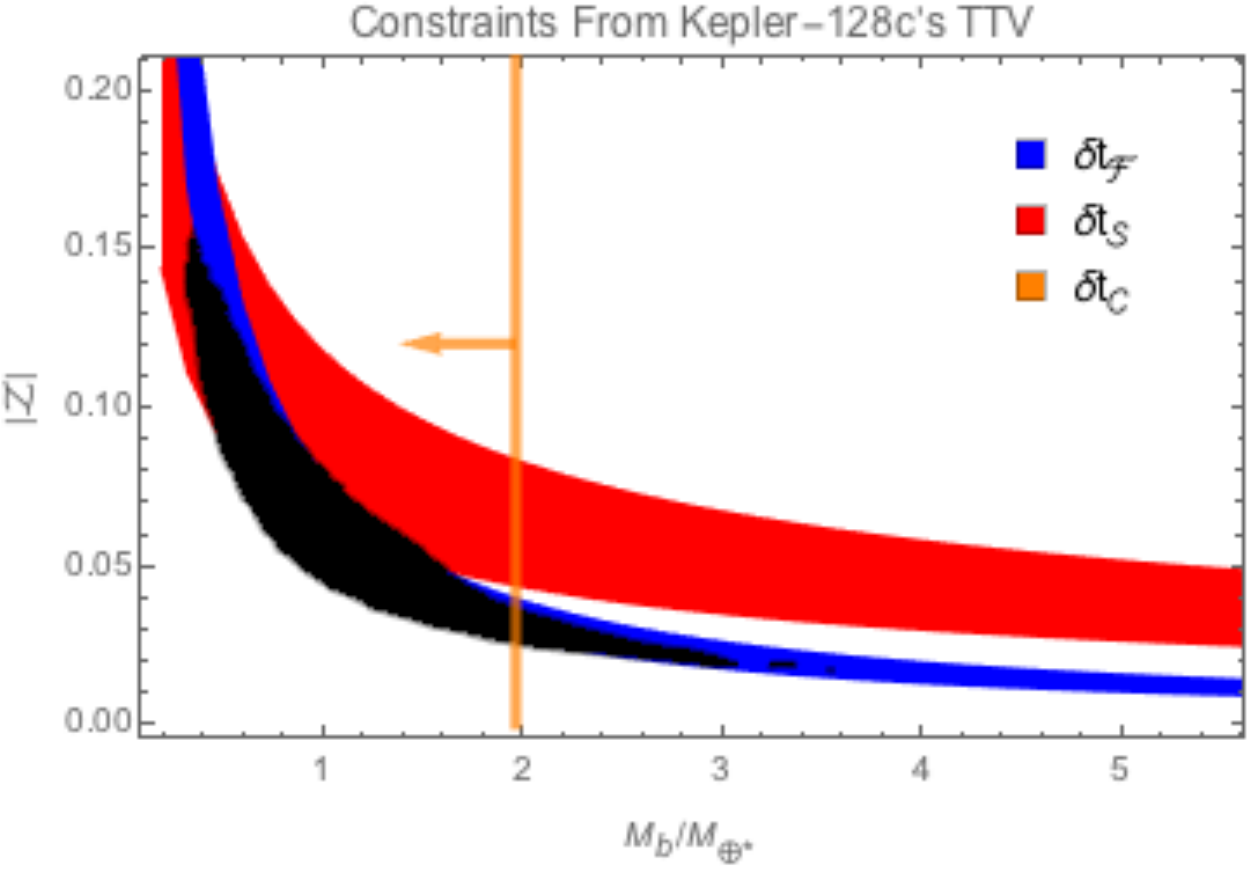}
	\caption{
	Analytic Constraint Plots for Kepler-128 (see Figure \ref{fig:kep307constraints} for description).
	The bottom panel shows a 1-$\sigma$ mass upper bound derived from the (lack of) chopping TTV amplitude.  
}
	\label{fig:kep128constraints}
	\end{center}
\end{figure}

Figure \ref{fig:kep128constraints} shows the analytic constraint plots (Section \ref{sec:methods}) for the inner and outer planets.
The TTVs of both  Kepler-128b and c possess non-zero  secondary components in addition to strong fundamental signals.
The results of the N-body MCMC are largely contained within the intersections of the constraints derived from these components. 
Figure \ref{fig:kep128constraints} shows that the fundamental and secondary TTV signals mainly place upper limits on $|\cal Z|$ or, equivalently, lower limits on masses.
The MCMC posteriors  possess long high mass tails that reflect the lack any strong upper limits from components of the TTV (Figure \ref{fig:kep128mass}).
The lower mass limits from MCMC and the analytic constraints indicate that both planets most likely have densities $\gtrsim 3$ g/cm$^3$ (Figure \ref{fig:MassRadius}).
The TTVs do not provide strong upper limits on the planet masses. 
The TTV of Kepler-128c provides a modest 1-$\sigma$ upper limit of $M_b<2.0~M_{\oplus*}$,
However, at the 2-$\sigma$ confidence level, this upper limit is extended to $M_b< 31~M_{\oplus*}$.

One can derive an upper limit on planet masses by requiring that Kepler-128b and c have physically plausible bulk densities. 
Pure iron planets with the same radii as Kepler-128b and c would have a masses of $\sim 4M_\oplus$ according to the models of \citet{Fortney:2007hf}.
Imposing an maximum mass of $4M_\oplus$ on Kepler-128b and c requires eccentricities of  $|{\cal Z}| \gtrsim 0.02$ based on the fundamental TTV amplitudes (see Figure \ref{fig:kep128constraints}).

\subsection{Kepler-26b and c (KOI-250)}
\label{sec:kep26}
\begin{figure}[htbp]
\begin{center}
\includegraphics[width=0.47\textwidth]{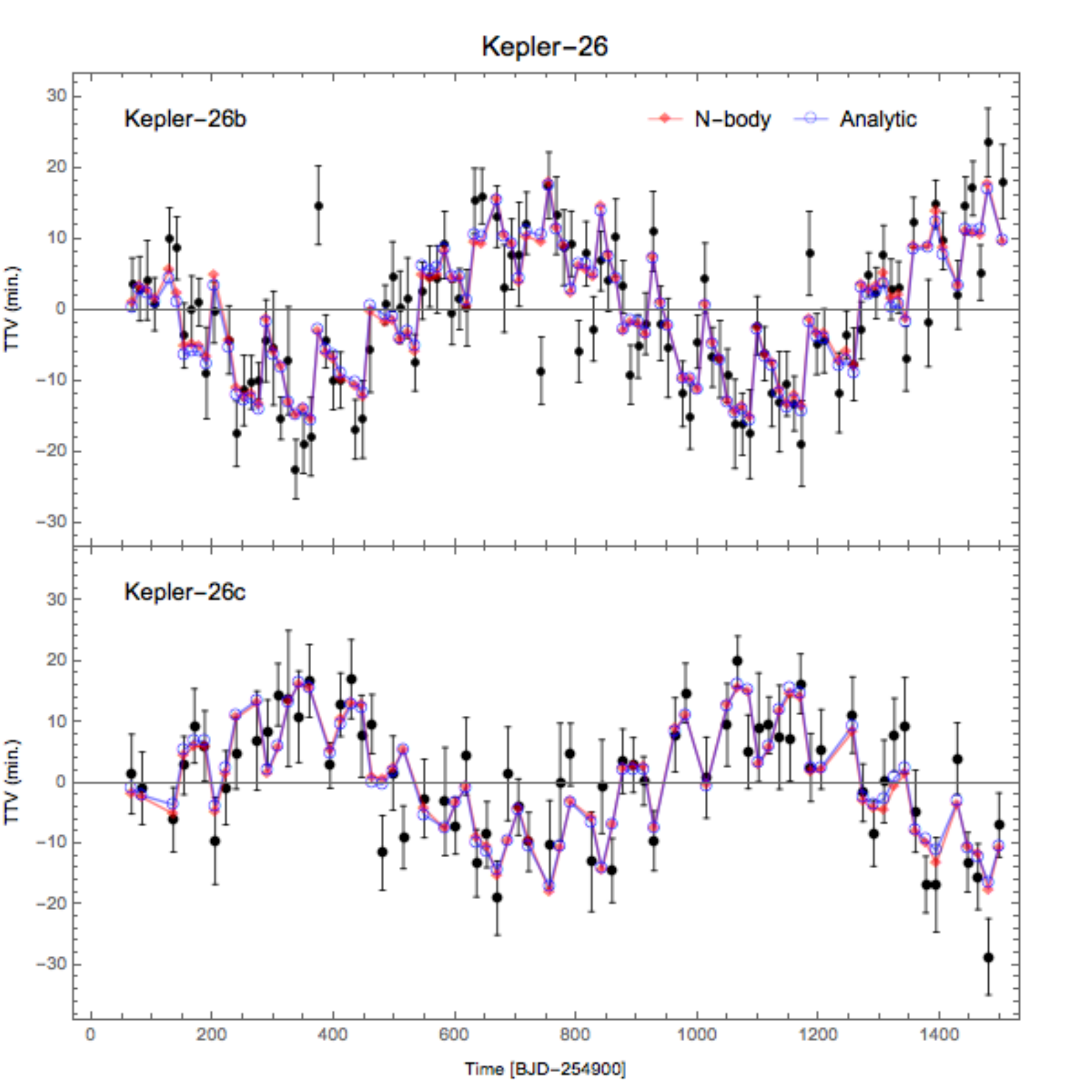}
\caption{ 
TTVs of the Kepler-26 system (see Figure \ref{fig:kep307ttv} for description).
}
\label{fig:kep26ttv}
\end{center}
\end{figure}

 Kepler-26b and c are a pair of sub-Neptune sized planets near the second order 7:5 MMR. 
 The planets were first confirmed by \citet{2012MNRAS.421.2342S} on the basis of anti-correlated TTVs.
 Both planets' TTVs, shown in Figure \ref{fig:kep26ttv}, show strong $\delta t_{\cal S}$ TTV amplitudes associated with their proximity to the 7:5 MMR as well as fast frequency chopping. 

\begin{figure}[htbp]
\begin{center}
\includegraphics[width=0.45\textwidth]{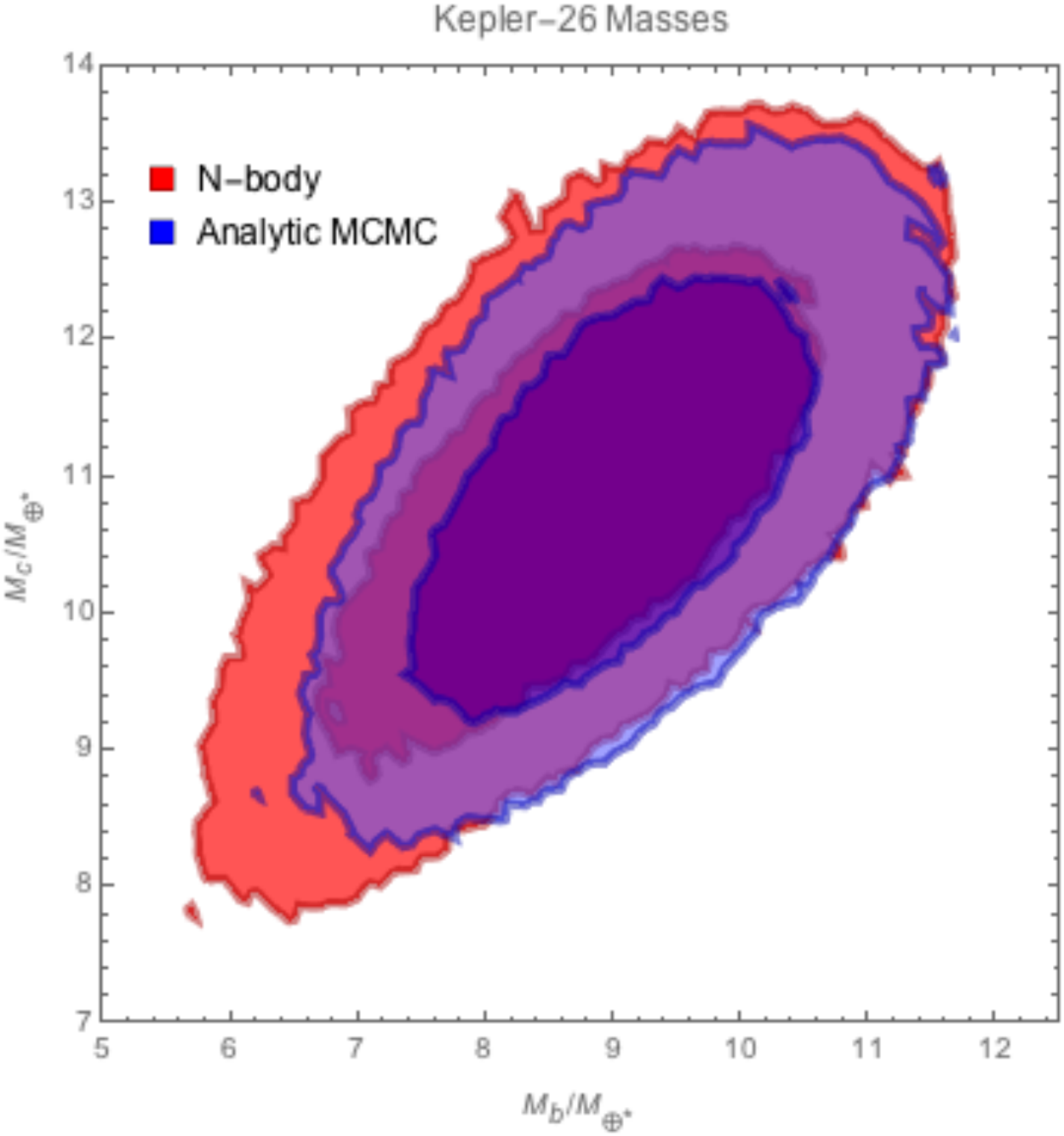}
\caption{
MCMC mass posterior for the Kepler-26 system (see Figure \ref{fig:kep307mass} for description).
}
\label{fig:kep26mass}
\end{center}
\end{figure}

\begin{figure}[htbp]
	\begin{center}
	\includegraphics[width=0.45\textwidth]{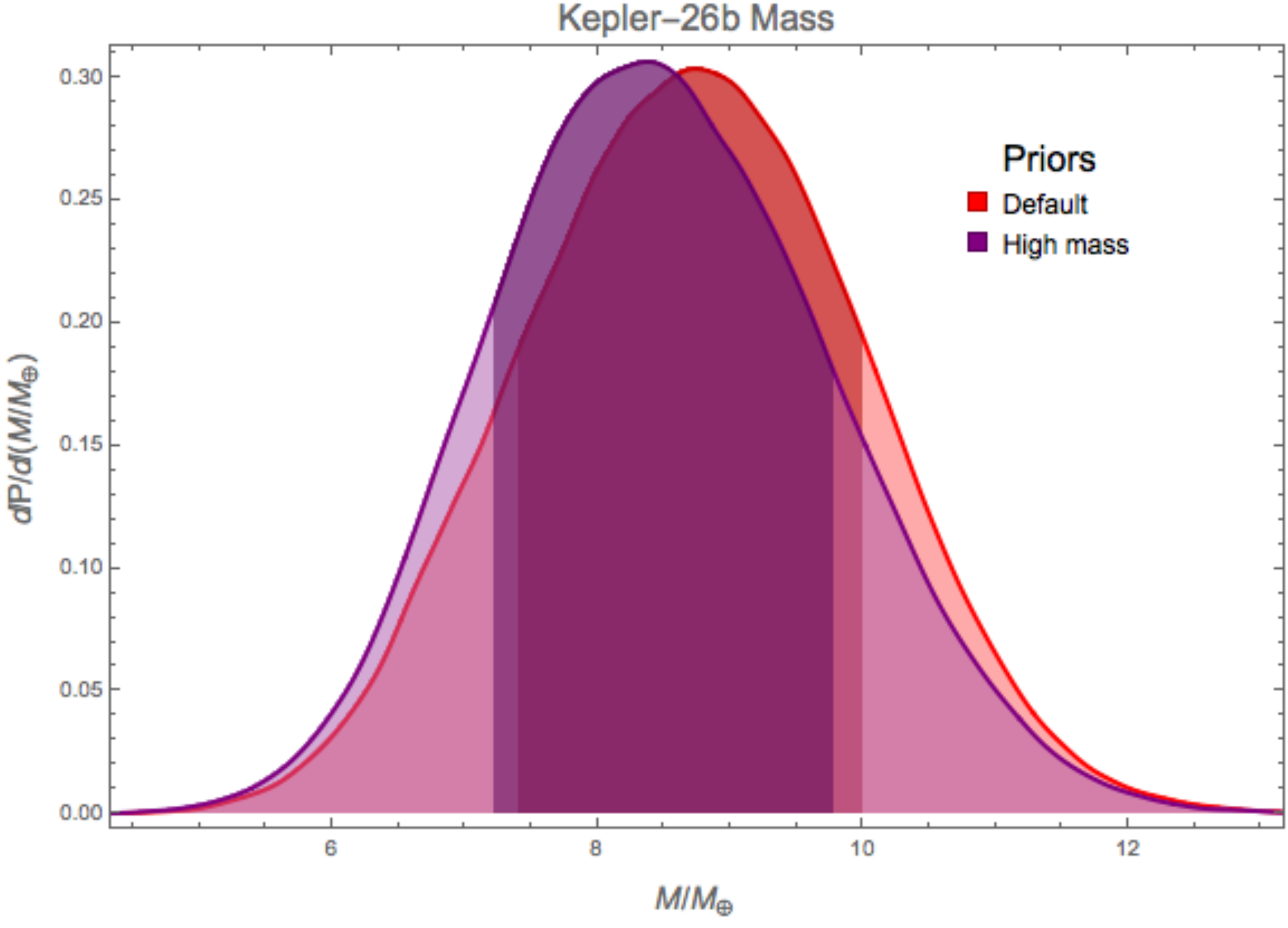}
	\includegraphics[width=0.45\textwidth]{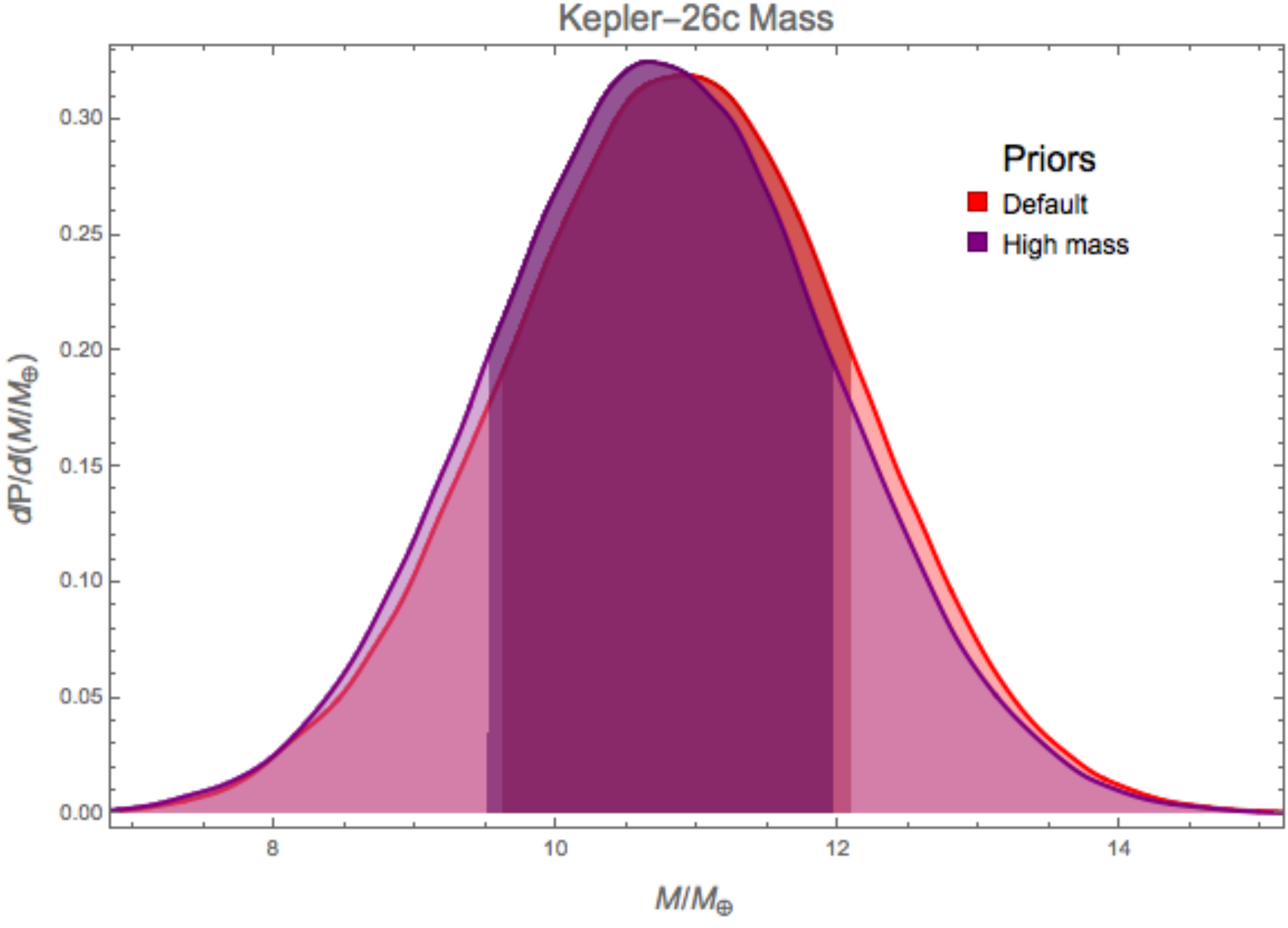}
\caption{
Comparison of MCMC priors for the Kepler-26 system  (see Figure \ref{fig:kep307mass_prior} for description).
}
	\label{fig:kep26mass_prior}
	\end{center}
\end{figure}

For the N-body MCMC an ensemble of 800 walkers was run 250,000 iterations, saving every 800th iteration.
The MCMC yielded $\sim 6,300$ independent posterior samples, based on analysis of the walkers' auto-correlation lengths.  
Joint mass constraints for Kepler-26b and c derived from both the N-body and analytic MCMCs are plotted in Figure \ref{fig:kep26mass}. 
The analytic and N-body MCMC results show good agreement. 
Figure \ref{fig:kep26mass_prior} shows that the inferred planet masses are essentially unaffected by adopting the alternate `high mass' priors (see Section \ref{sec:methods}). 

\begin{figure}
\begin{center}
\includegraphics[width=0.45\textwidth]{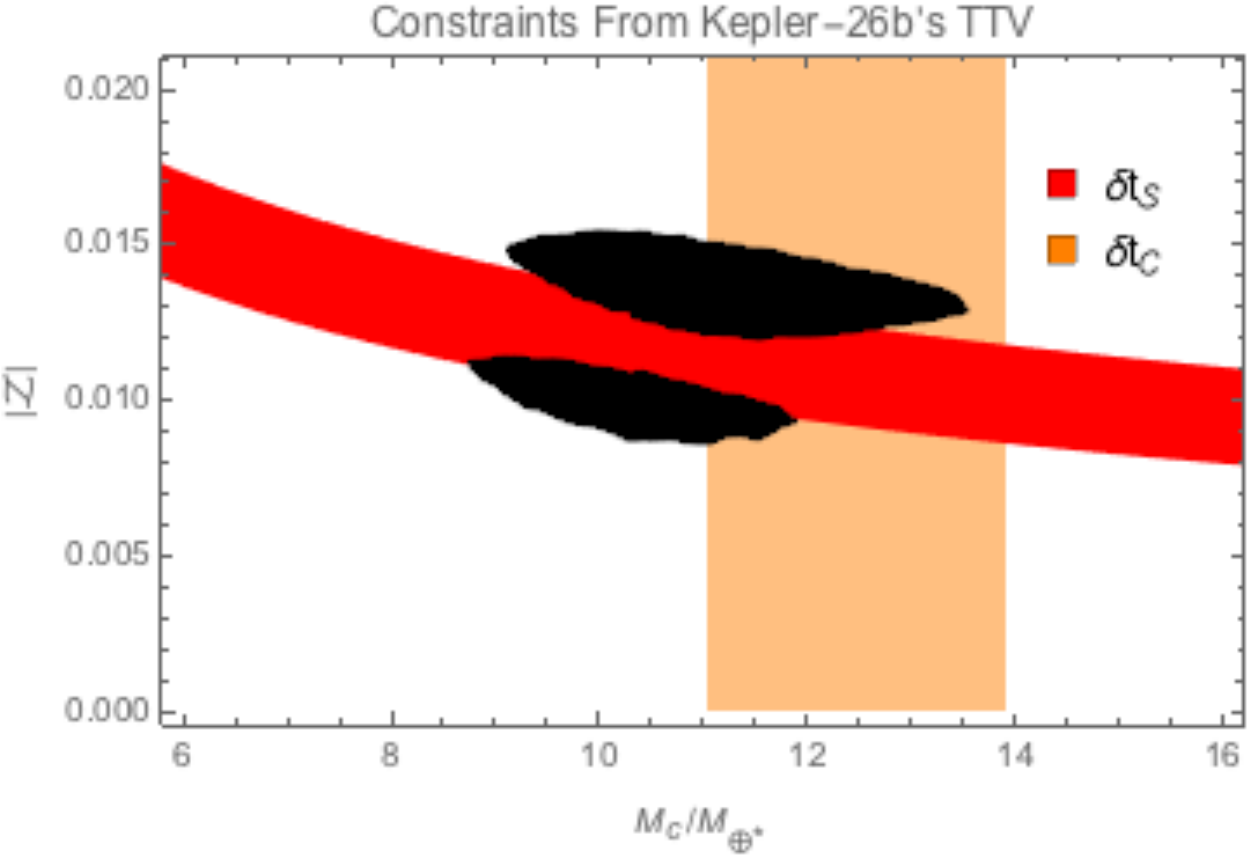}
\includegraphics[width=0.45\textwidth]{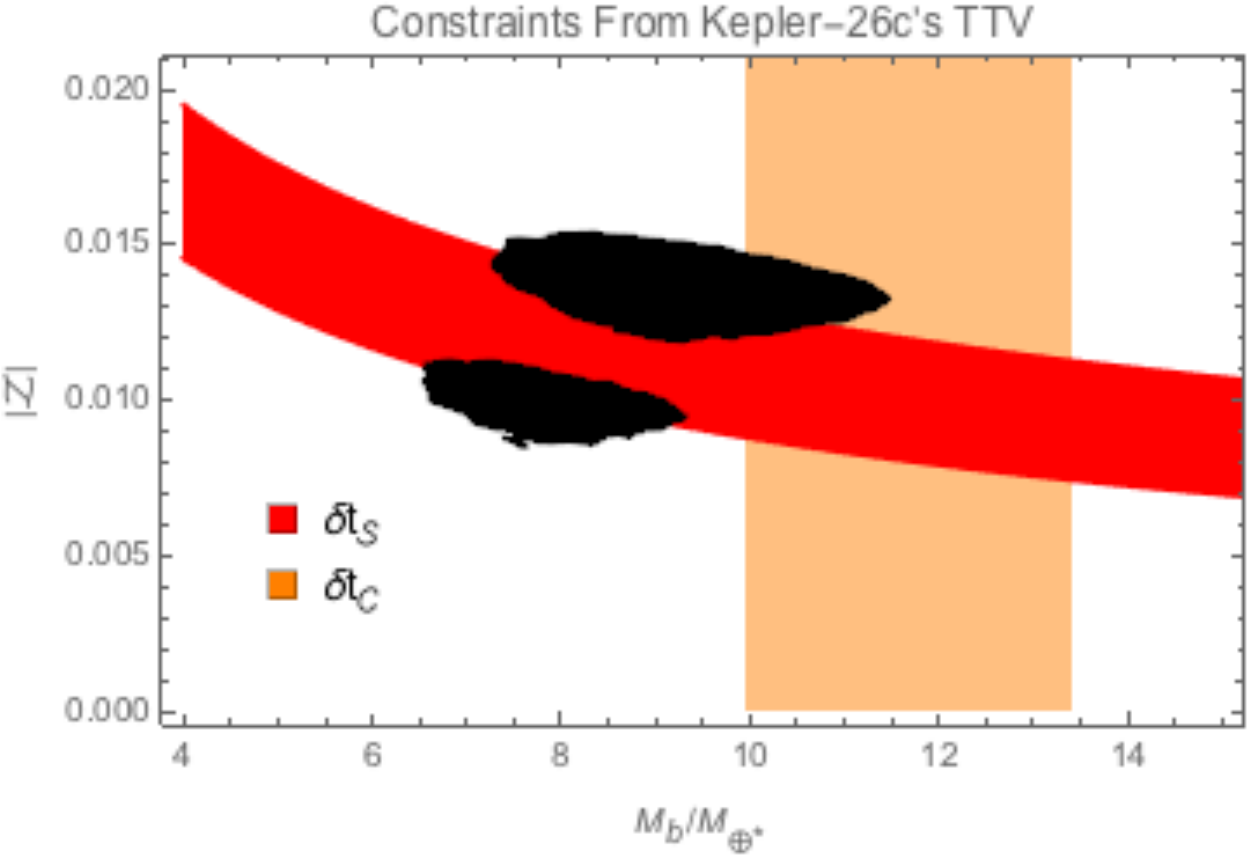}
\caption{
	Analytic Constraint Plots for Kepler-26 (see Figure \ref{fig:kep307constraints} for description).
}
\label{fig:kep26constraints}
\end{center}
\end{figure}

Figure \ref{fig:kep26constraints} shows the analytic constraints plot for both planets.  
The combined constraints from the  7:5 MMR $\delta t_{\cal S}$ and chopping signals roughly agree with the MCMC results.
The MCMC results plotted in Figure \ref{fig:kep26constraints} show that the posterior is bimodal.
This bimodality is expected: $\delta t_{\cal S}$ is a quadratic polynomial in $\cal Z^*$ and so for any value of $\mu$ there are two roots for $\cal Z^*$ that give the same $\delta t_{\cal S}$ signal.

The Kepler-26 system hosts two additional confirmed planets, Kepler-26d and e.  The periods of these two planets, $P_d=3.5$ days and $P_e=46.8$ days, place them far from planets `b' and `c'  and they are unlikely to have an appreciable influence the TTVs of `b' and `c' given their sizes, $R_d\sim1.1~R_\oplus$ and $R_e\sim2.4~R_\oplus$.
	
\subsection{Kepler-33 (KOI-707)}
\label{sec:kep33}
\begin{figure}
\begin{center}
\includegraphics[width=0.45\textwidth]{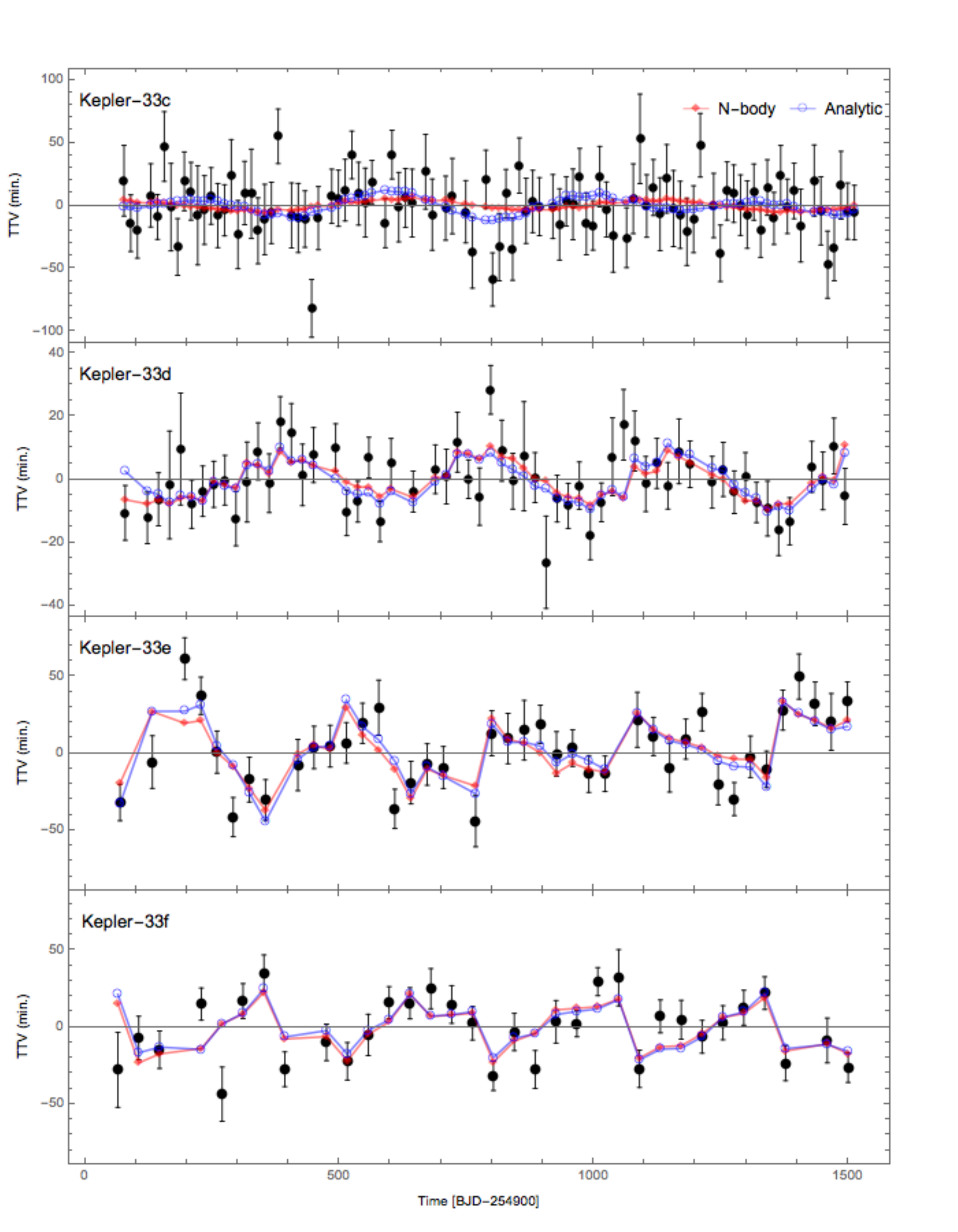}
\caption{ 
TTVs of the Kepler-33 system (see Figure \ref{fig:kep307ttv} for description).
}
\label{fig:kep33ttv}
\end{center}
\end{figure}

	Kepler-33 hosts 5 planets confirmed by \citet{2012ApJ...750..112L} ranging in size from $\sim 1.7~R_\oplus$ to $\sim 5.3~R_\oplus$ . We model only the TTVs of the outer 4 planets, ignoring the innermost planet, 
    Kepler-33 b.\footnote{Kepler-33b has a period of $5.67$ days and a radius of $1.7 \pm 0.18 R_\oplus$. The relative distance of Kepler-33b from any low order mean motion resonances with the other planets combined with its small size imply its influence on their TTVs should negligible.}
The outer four planets are arranged in a closely packed configuration near a number of first and second order MMRs. Planets `c' and `d' lie near the second-order 5:3 MMR ($\Delta=-0.008$). Planets `d' and `e' lie near a 3:2 MMR ($\Delta=-0.027$) and the pair `e' and `f' are close to the 9:7 MMR ($\Delta=0.004$) and fall between the 4:3 and 5:4 MMRs ($\Delta = -0.032$ and $+0.032$, respectively). This configuration also places planets `d' and `f' somewhat near the 2:1 MMR with $\Delta_2=-0.058$. Figure \ref{fig:kep33ttv} shows the TTVs of Kepler-33 and the best-fit N-body and analytic models.

\begin{figure}[htbp]
	\begin{center}
	\includegraphics[width=0.47\textwidth]{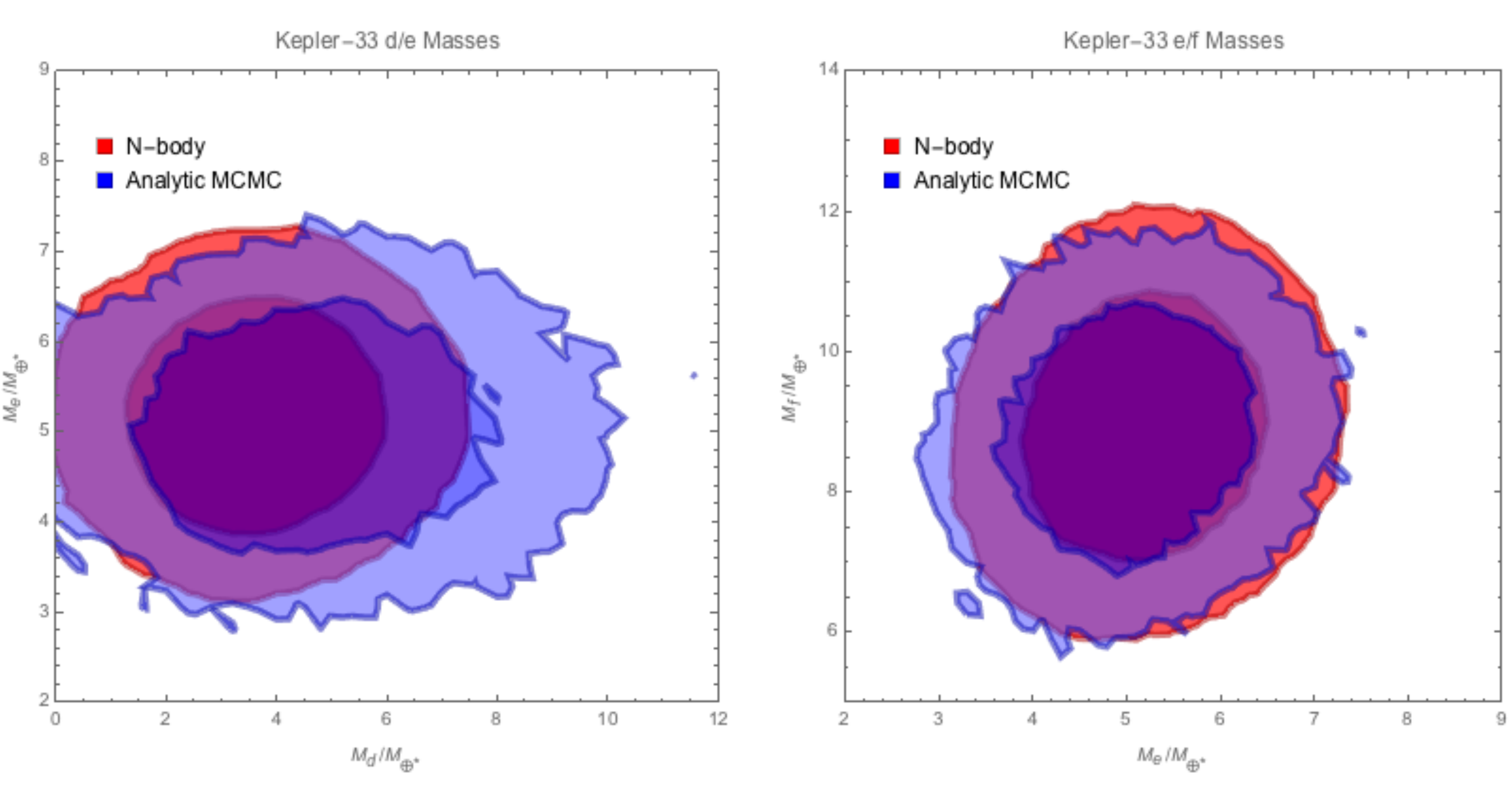}
\caption{ 
MCMC mass posterior for the Kepler-33d, e, and f (see Figure \ref{fig:kep307mass} for description).}
	\label{fig:kep33mass}
	\end{center}
\end{figure}

\begin{figure}[htbp]
	\begin{center}
	\includegraphics[width=0.45\textwidth]{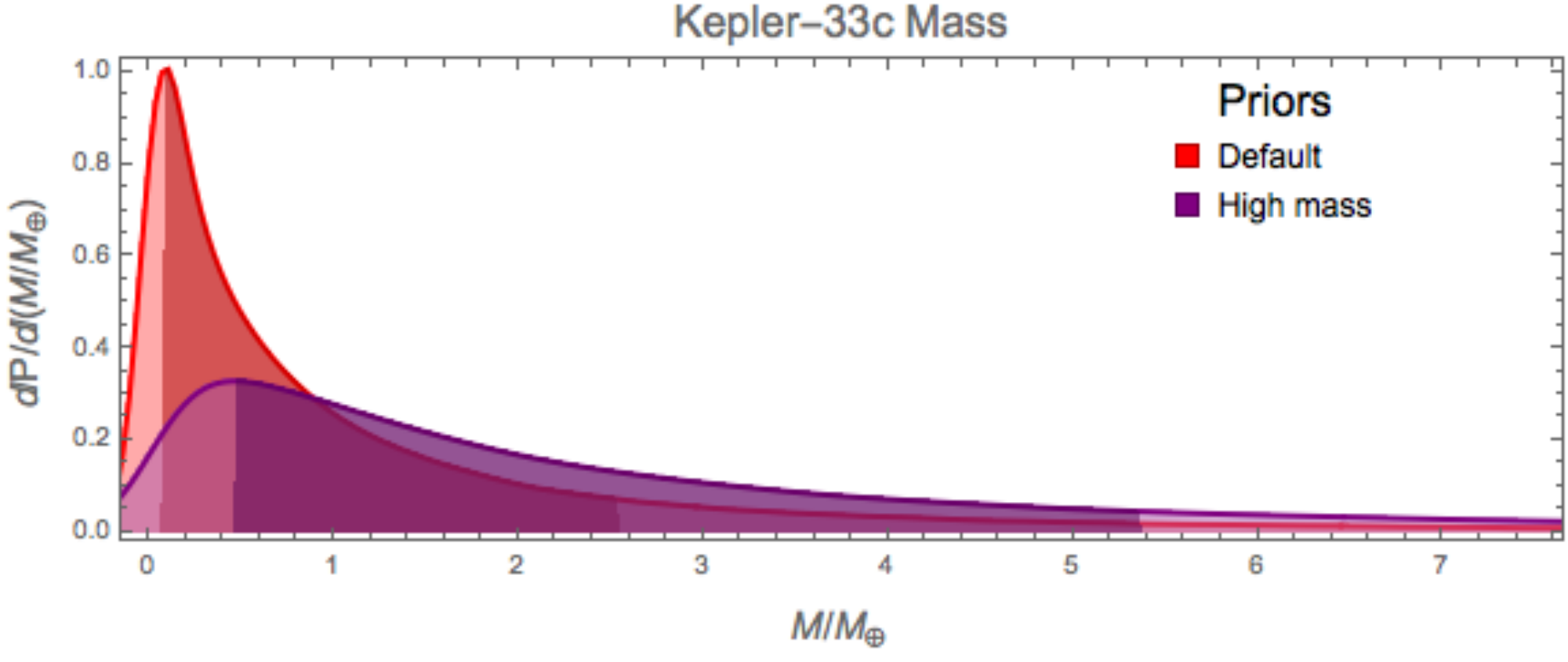}
	\includegraphics[width=0.45\textwidth]{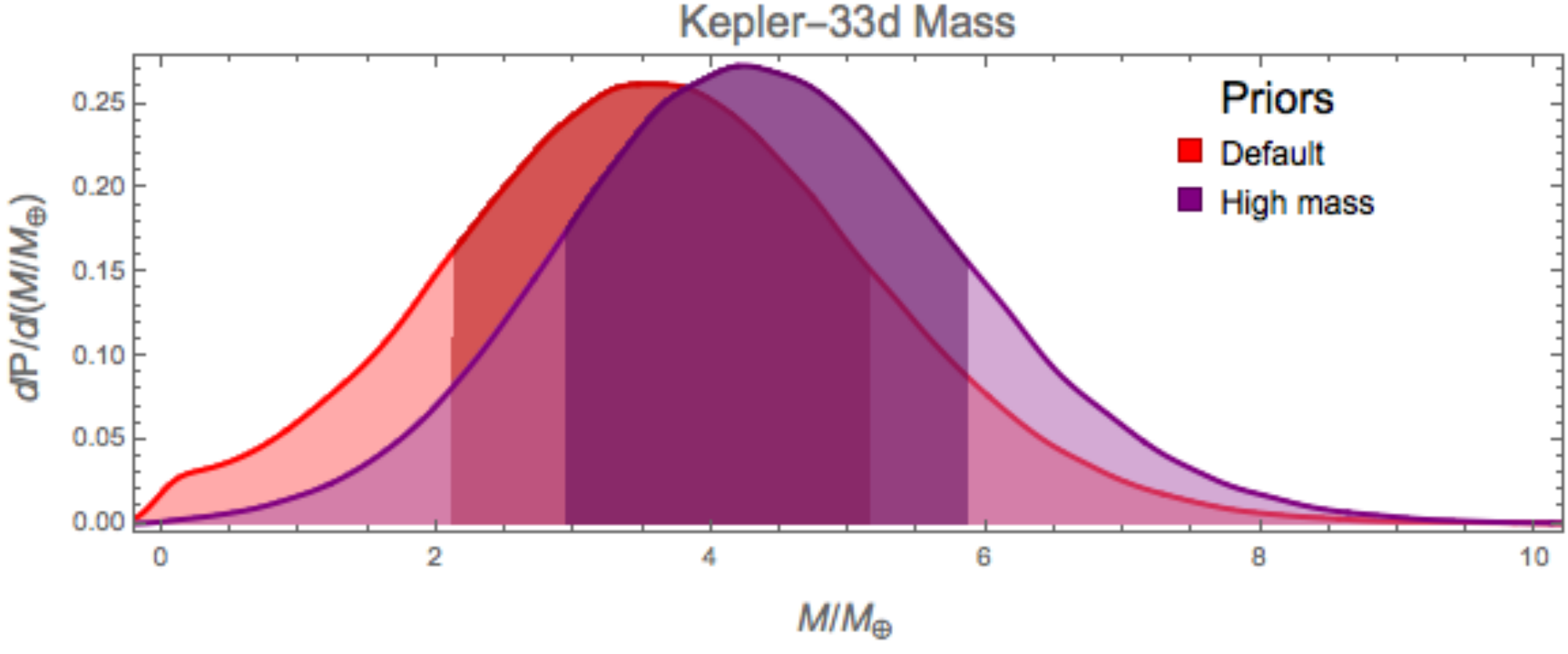}
	\includegraphics[width=0.45\textwidth]{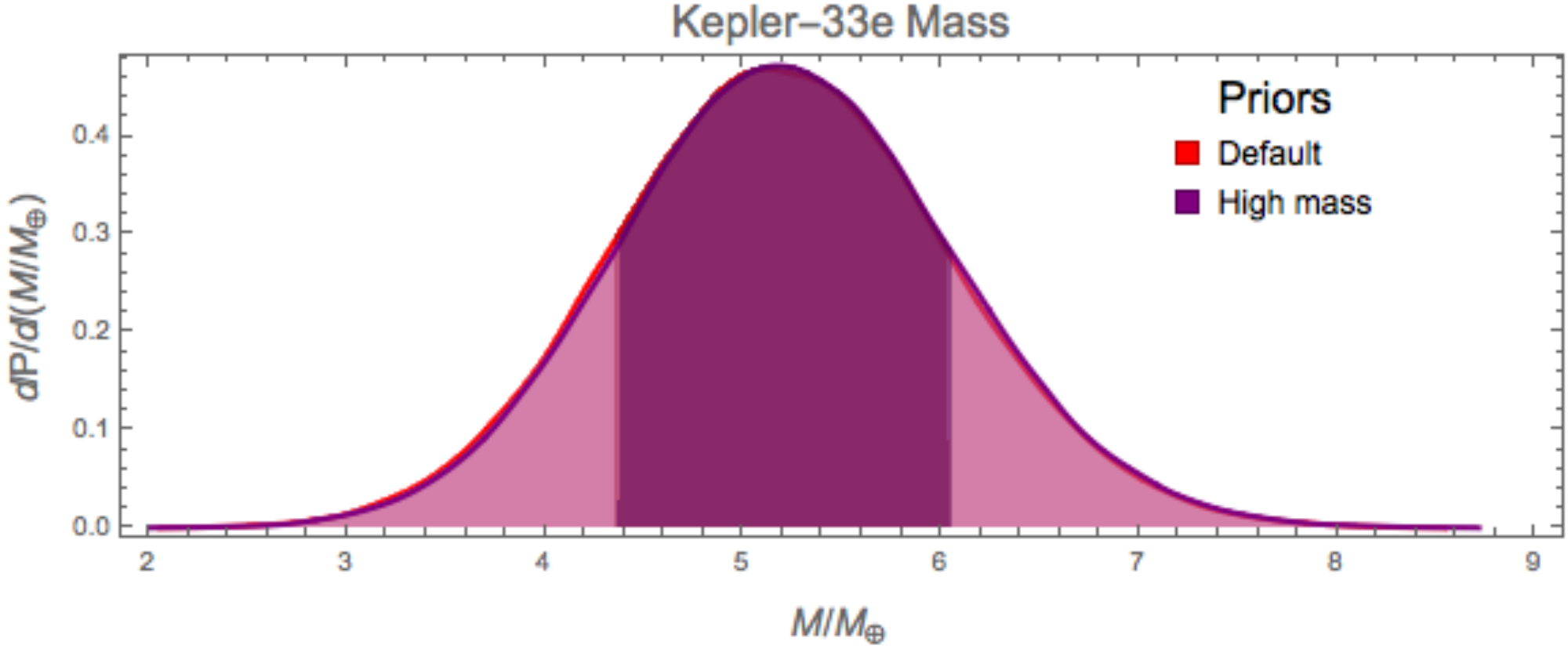}
	\includegraphics[width=0.45\textwidth]{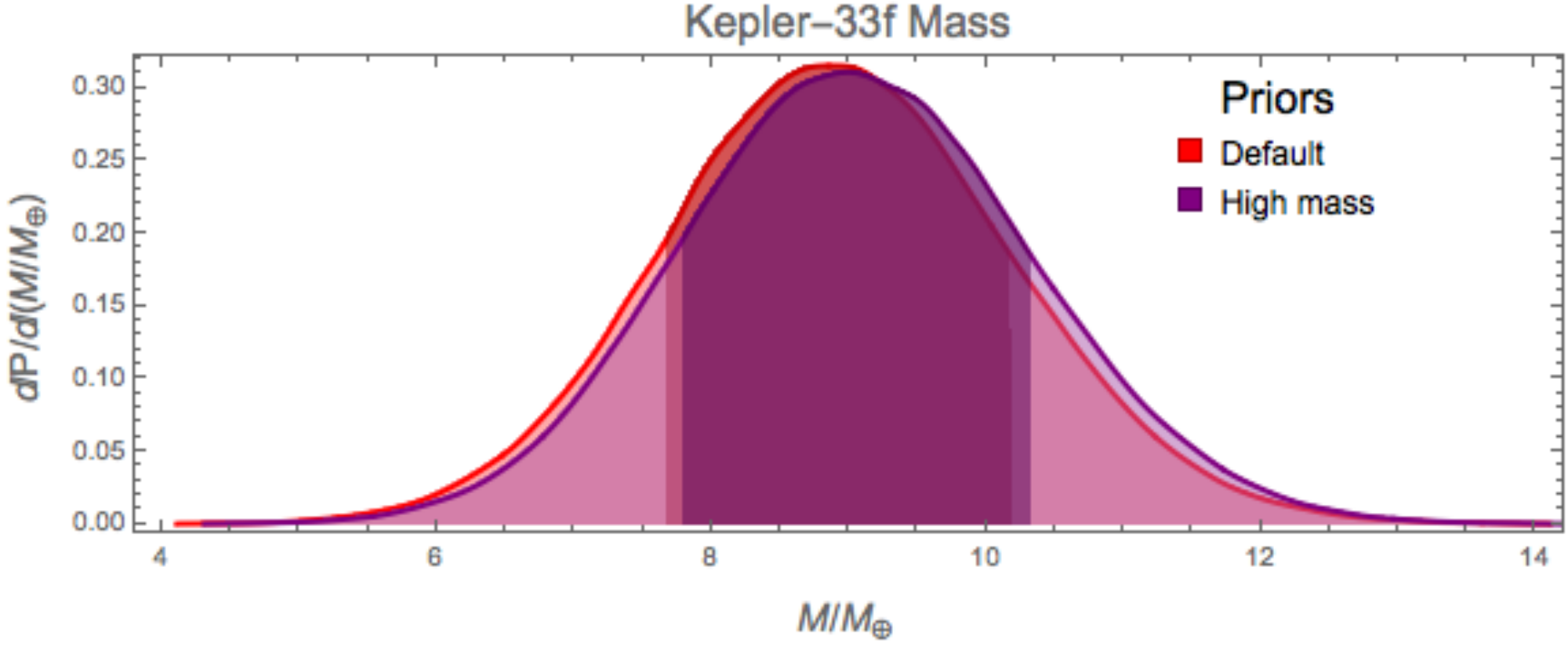}
\caption{
Comparison of MCMC priors for the Kepler-33 system  (see Figure \ref{fig:kep307mass_prior} for description).}
	\label{fig:kep33mass_prior}
	\end{center}
\end{figure}

 For the N-body MCMC, an ensemble of 1000 walkers were evolved for 300,000 iterations, saving every 800th iteration.
 This resulted in $\sim 25,200$ independent posterior samples based on analysis of the walker auto-correlation lengths. 
 The planet mass constraints derived from MCMC for planets `d',`e', and `f'  are plotted in Figure \ref{fig:kep33mass}.  
 The mass of the innermost planet, Kepler-33c, is poorly constrained, with the MCMC mainly providing an upper limit  (see Figure \ref{fig:kep33mass_prior}).
 Figure \ref{fig:kep33mass_prior} compares MCMC results using default and high mass priors. 
 The inferred masses of planets `e' and `f' are nearly unaffected  by the choice of prior. 
 The inferred mass of planet `c' and, to a lesser extent, `d' are sensitive to the assumed prior, indicating that these planets' masses are not as constrained by the transit time data.

\begin{figure}
\begin{center}
\includegraphics[width=0.45\textwidth]{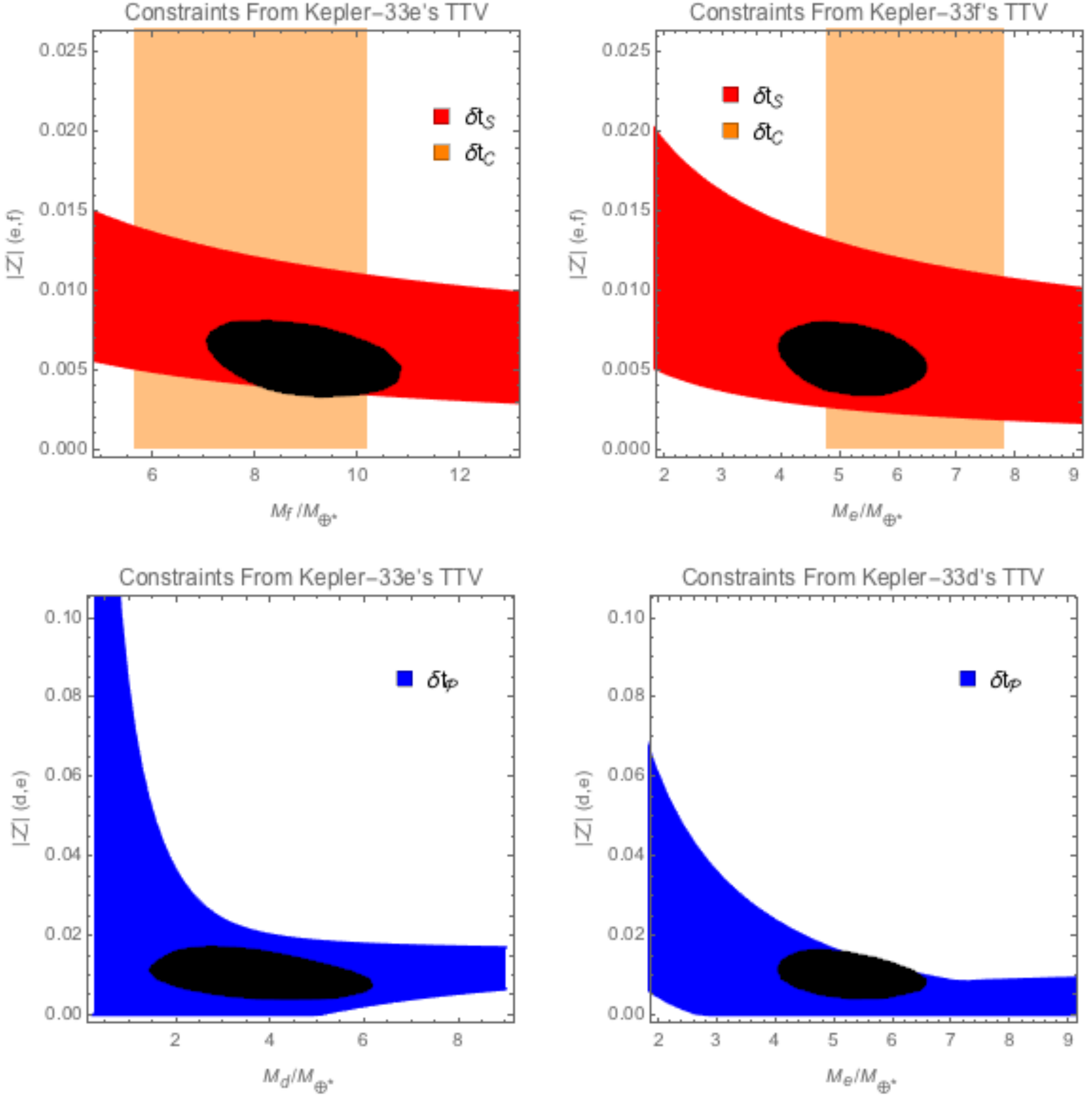}
\caption{
	Analytic Constraint Plots for Kepler-33: The top left panel shows the constraints provided by 
	planet `e''s secondary (9:7) and chopping TTV amplitudes caused by planet `f' (see Section \ref{sec:analytic_model}). 
	The top right panel shows the constraints provided by the secondary and chopping amplitudes in the TTV of 
	planet `f' caused by planet `e'.
	The bottom left (resp. right) panel shows the constraints provided by the fundamental TTV amplitude of planet `e' (resp. `d')
	caused by planet `d' (resp. `e').
	The black regions shows the N-body MCMC result, at 68\% confidence. The bottom panel 
	shows the same for the outer planet.  
}
\label{fig:kep33constraints}
\end{center}
\end{figure}

Analytic constraint plots for the  Kepler-33 system are shown in Figure \ref{fig:kep33constraints}.
The top row shows the masses and $|{\cal Z}|$ of planets `e' and `f'.
The MCMC results for planet `e' and are `f' are explained well by the joint constraints derived from their mutual chopping and 9:7 $\delta t_{\cal S}$  signals.
The MCMC  constraints for planet `d' and `e' are consistent with the constraint derived from their 3:2 fundamental TTV signals. 
The masses and ${\cal Z}$ of planet `d' and `e' would be degenerate based solely on the observed $\delta t_{\cal F}$ signals.
However, the mass of planet `e' is already constrained by interactions with planet `f'. 
Since the mass of planet `e' is constrained, the combined eccentricity, ${\cal Z}$, of planet `d' and `e' can be inferred from the fundamental signal in the TTV of planet `d'. 
With ${\cal Z}$ constrained by the planet `d' fundamental TTV, the mass of planet `d' is in turn constrained by the fundamental TTV signal it induces in planet `e'.

 \section{Summary and Discussion}
  \label{sec:summary}
 We have presented an analytic model for the TTVs of multi-planet systems and conducted N-body MCMC simulations to infer planet properties. 
The analytic constraints show good agreement with N-body fits and provide an clear explanation of the MCMC results.   
We also demonstrate that the planet masses derived from MCMC are insensitive to the assumed priors. 
   We summarize the key features of our analytic model:
\begin{enumerate}
\item  We derive an anlytic treatment of the influence of second-order MMRs on TTVs.  The effects of second-order MMRs can help to constrain planet masses and eccentricities both near first-order resonances, as in the case of Kepler-128 (Section \ref{sec:kep0274}); or planets near a second-order resonance such as Kepler-26 (Section \ref{sec:kep26})
\item We identify the combined eccentricity, $\cal Z$, as a key parameter in determining the TTV signal.  
Eccentricities of individual planets will rarely be constrained from TTVs alone.
Extracting $|{\cal Z}|$ from N-body fits provides a useful way to interpret the results.
\item  The analytic constraint plots show that a simple linear least-squares fit can be used to derive approximate constraints from TTVs with minimal computational burden. 
\end{enumerate}

With the exception of the Kepler-128 system, the planets have low densities, likely less dense than water (Figure \ref{fig:MassRadius}). 
These planets are new additions to the growing ranks of low-density sub-Neptune sized planets that have been characterized via TTV observations. 
The density uncertainties for two of the systems,  Kepler-26 and Kepler-307, are dominated by uncertain planet radii (Figure \ref{fig:MassRadius}).
	
The combined eccentricities are small (Table \ref{tab:ecc}), as expected from previous work on the eccentricities of TTV systems \citep{Wu:2013cp,2014ApJ...787...80H}.
{\it In situ} formation scenarios with merging collisions predict substantially larger eccentricities  \citep[$e\sim0.1$, ][]{2013ApJ...775...53H}.

In future work we plan to apply the techniques developed in this paper to more systems.

\acknowledgments{
  Acknowledgments. We thank Ben Farr, Tyson Littenberg, and Jason Steffen for helpful discussions.
  We are also grateful to the Kepler team for acquiring and publicly releasing such spectacular results.
  This research has made use of the NASA Exoplanet Archive, which is operated by the California Institute of Technology,
   under contract with the National Aeronautics and Space Administration under the Exoplanet Exploration Program.
 SH acknowledges support from the NASA Earth and Space Science Fellowship program, grant number NNX15AT51H. 
 YL acknowledges grants AST-1109776 and AST-1352369 from NSF, and NNX14AD21G from NASA.
}

\bibliographystyle{apj}
\bibliography{Untitled}
\clearpage

\appendix

\section{A: analytic TTV Formulae}
\label{sec:appendix}
We derive the analytic TTV formulae for two  interacting coplanar planets, working to leading  order in the planet-star mass ratio ($\mu\equiv m/M_*\ll 1$) and assuming that the eccentricities ($e$) are small.  In particular, we drop all terms that are $O(e^3)$ and higher and only retain terms that are $O(e^1)$ or $O(e^2)$ when they are accompanied by resonant denominators. Our formulae are meant to apply to the bulk of {\it Kepler} planets, but they will fail for planets  close to a third- (or higher-) order MMR, or if the planets are librating in resonance.
	
We start with a detailed derivation of the case of a planet perturbed by an exterior companion, 
the results of which are in  \ref{sec:expand}.  

\subsection{\label{sec:derivation}\ref{sec:derivation}: Derivation (External Perturber)}
Our notation mostly follows \citet{1999ssd..book.....M}. In particular, primed/unprimed variables refer to the outer/inner planet, ${\bf r}$ is the (astrocentric) position vector;   $a, \lambda, e, \varpi$ are the (astrocentric) semimajor axis,  mean longitude, eccentricity, and longitude of pericenter; and $\alpha=a/a'$.   Note,  however, that we use $\mu$ for $m/M_*$, whereas  Murray \& Dermott use it for $Gm$.

\subsubsection{\label{sec:ztot} \ref{sec:ztot}: From orbital elements ($\delta z$ and $\delta \lambda$) to TTV ($\delta t$)}

The angular position of a planet relative to the line of sight is $\theta$; it is related to the  orbital elements via $\theta= \lambda + 2e\sin(\theta - \varpi)+O(e^2)$. 
 It will prove convenient to replace the elements $e$ and $\varpi$ with the complex eccentricity $z$ \citep{2007MNRAS.374..131O}:
\begin{eqnarray} 
	z&\equiv& ee^{i\varpi} 
\end{eqnarray}
implying
\begin{eqnarray}
	\theta &=& \lambda + \left({z^*\over i} e^{i\theta} +c.c.\right)
\label{eq:ThetaToOrbel}
\end{eqnarray}
where ``$c.c.$'' means the complex conjugate of the preceding term, and we drop  $O(e^2)$ terms because they are unaccompanied by any resonant denominators. We expand  the orbital elements into their unperturbed Keplerian values plus perturbations due to the companion that are $O(\mu')$:
\begin{eqnarray*}
	\label{eq:a_plus_da}
	a(t) &= a_0 + \delta a(t) \\
	\label{eq:z_plus_dz}
	z(t) &= z_0 + \delta z(t)\\
	\label{eq:l_plus_dl}
	\lambda(t) &= \lambda_0 + \delta\lambda(t) 
\end{eqnarray*}
where $a_0$ and $z_0$ are constant, and
\begin{eqnarray*}
	\lambda_0 = n(t-T) = {2\pi \over P}(t-T) \ ,
\end{eqnarray*}
expressed in terms of the  constants $n, P$, and $T$---which are respectively the mean motion, orbital period, and reference time. We write the times of transit as $t_{\rm trans,0}+\delta t$, where  $\delta t$  is the TTV; i.e., it is the $O(\mu')$ perturbation in the transit time due to the companion. Setting $\theta=0$ in Equation \eqref{eq:ThetaToOrbel} then implies at $O(\mu')$ \citep[][]{2008ApJ...688..636N}:
\begin{equation}
	\delta t  = -\frac{P}{2\pi}\left( \delta\lambda +\left(\frac{\delta z^*}{i} + \text{c.c.}\right)\right)
	\label{eq:ttv_form}
\end{equation}

\subsubsection{\label{sec:eom}\ref{sec:eom}: Equations of motion}

 We shall solve perturbatively for  $\delta\lambda$ and $\delta z$, which then give the TTV via Equation \eqref{eq:ttv_form}. The equations of motion for our preferred variables, $\{a,z,\lambda\}$, are Hamilton's equations for the corresponding canonical variables \citep{2007MNRAS.374..131O}:
\begin{align}
		&\frac{dz}{dt} = 2i n'\frac{\mu'}{\sqrt{\alpha}} \pd{R}{z^*}   \label{eq:zdot} \\
		&\frac{d\ln a}{dt} = 2n'\frac{\mu'}{\sqrt{\alpha}} \pd{R}{\lambda}  \label{eq:adot} \\
		&\frac{d\lambda}{dt} = \frac{n'}{\alpha^{3/2}}\left(1-\frac{3}{2} \frac{\delta a}{a}\right)-2 n' \mu'\sqrt{\alpha}{\partial{R}\over{\partial\alpha}} \label{eq:ldot}
\end{align}
where the bracketed term in $d\lambda/dt$ comes from the partial derivative of the Keplerian Hamiltonian, expanded to first order in $\mu'$ and we have dropped terms that are smaller by a factor  $O(e^2)$\footnote{We have dropped a term from the right-hand side of Equation (\ref{eq:ldot}). In truth, one should  replace $\partial R/\partial a\rightarrow \partial R/\partial a-{z^*\over 4a}\partial R/\partial z^*$. However, that term does not contribute to the TTV to the order of approximation at which we work. More precisely, its contribution near a $J$:$J$-1 MMR is suppressed by the large factor $n_{J,1-J}$, (see Section \ref{sec:expand}).}. 
The disturbing function is
\begin{eqnarray}
R\equiv
\frac{a'}{|{\bf r} - {\bf r'}|} - a'\frac{{\bf r} \cdot {\bf r'}}{|{\bf r'}|^3} \label{eq:dist_func}
= 	\sum_{j,k}R_{j,k}e^{i (j\lambda' + k \lambda) }	\label{eq:R_Form}
\end{eqnarray}
for which the Fourier amplitudes $R_{j,k}$  are 
 given in \cite{1999ssd..book.....M}.  For our purposes, the following terms  suffice up to $O(e^2)$:
\begin{eqnarray}
		R_{j,-j} &=& f_1^j - \delta_{j,1}\frac{\alpha}{2} \label{eq:R_zeroth}\\
		R_{j,1-j} &=& \frac{1}{2}\left((f_{27}^j +\delta_{j,1}\frac{3}{2}\alpha-\delta_{j,-1}\frac{1}{2}\alpha)z^* + (f_{31}^j-\delta_{j,2} 2\alpha) z'^*\right) \label{eq:R_first} \\
		R_{j,2-j} &=&\frac{1}{2}\left( f_{45}^j z^{*2} +f_{49}^jz^*z'^* +\left( f_{53}^j -\delta_{j,3}\frac{27\alpha}{8}\right) z'^{*2}\right) \label{eq:R_second}
\end{eqnarray}
where 
 the $f^{j}_N$  are combinations of Laplace coefficients and their derivatives  whose explicit form is listed in the Appendix of \citet{1999ssd..book.....M}.\footnote{ We omit indirect terms  with $j$=-1,1, and 2 in  Equation \eqref{eq:R_second} because they will never appear with small denominators in Equations \eqref{eq:tS_1} or \eqref{eq:t2j} below and can therefore be ignored for our purposes.}
	Our $R$ is related to Murray \& Dermott's ${\cal R}$ via $R =(a^\prime/Gm){\cal R}$.


\subsubsection{\label{sec:zsol}\ref{sec:zsol} Solutions for $\delta z$ and $\delta\lambda$}

The equations of motion  are  integrated by (a) replacing $\lambda$ in the exponentials with $\lambda_0$ (after taking the derivative  $\partial R/\partial \lambda$), 
which is valid to $O(\mu')$, and (b) matching Fourier coefficients. The result is
\begin{eqnarray*}
	\label{eq:z_Fourier}
	\delta z = \sum_{j,k}{z}_{j,k} e^{ i(j\lambda'_0 + k\lambda_0)} \ , \  \ 
	\label{eq:l_Fourier}
	\delta \lambda =  \sum_{j,k}   \lambda_{j,k} e^{i(j\lambda'_0 + k\lambda_0)} 
\end{eqnarray*}
where
\begin{align}
	\label{eq:zhat}
		{z}_{j,k} &= \frac{2\mu'}{\sqrt{\alpha}}n_{j,k}\frac{\partial R_{j,k}}{\partial z^*} \\
	\label{eq:lhat}
		{\lambda}_{j,k} &= \mu'\left(-\frac{3k}{\alpha^2 i} n_{j,k}^2  R_{j,k} - \frac{2}{i}n_{j,k} \sqrt{\alpha} 	\frac{\partial}{\partial \alpha} R_{j,k}\right) 
\end{align}
and we have defined
\begin{equation} 
	n_{j,k}\equiv {n'\over jn' + k n} \label{eq:njk}
\end{equation}
 Note that $n_{j,k}$ is related to $\Delta$, the fractional distance to the nearest first order $J$:$J$-1 MMR defined in the body of the paper (Eq. \ref{eq:delta_defn}), via
\begin{eqnarray}
n_{J,1-J} = -{1 \over{J \Delta}}
\label{eq:njdel}
\end{eqnarray}
and hence is large near resonance. 

The TTV is obtained by inserting $\delta z$ and $\delta \lambda$ into Equation \eqref{eq:ttv_form} and evaluating at the times of transit, i.e., setting $\lambda_0=0$ in the exponent.  We find
\begin{align}
 \delta t  &=\frac{P}{2 \pi i}\sum_{j> 0}e^{ij\lambda'_0} \left(\sum_k  z_{j,k}- z^*_{-j,-k}-i\lambda_{j,k}\right) + \text{c.c.} 
	\label{eq:TTV_Fourier_AliasedII}
\end{align}
To make $j>0$, we have rearranged terms, made use of the reality condition $(\lambda_{-j,-k}  =  \lambda^*_{j,k})$, and dropped the $j=0$ term because it does not contribute to the  TTV. Henceforth,  $j$ will be  restricted to positive values.


\subsection{\label{sec:expand}\ref{sec:expand}: 
 Explicit TTV Formulae (External Perturber)
}

We simplify Equation \eqref{eq:TTV_Fourier_AliasedII} by expanding up to second order in $e$:
\begin{equation}
\delta t =
\mu'{P\over 2\pi i}
 \sum_{j>0}e^{ij\lambda'_0}\left( t_j^{(0)}  +  t_j^{(1)}  + t_j^{(2)}\right) + \text{c.c.} \label{eq:dtAll}
\end{equation}
where  $t_j^{(m)} $ is $m$-th order in eccentricity. The time dependence enters only in the exponent ($\lambda_0'\propto n't+$ const.), and the $t_j^{(m)}$ depend on the $a_0$ and $z_0$ of the two planets. (Henceforth, we drop the subscript 0). We work out the three $t_j^{(m)}$ in turn.

\begin{itemize}
\item
 {\bf $t_j^{(0)}$:}  
 The amplitude of an $m$-th order MMR ($j$:$j$-$m$) is $m$-th order in eccentricity, i.e., $R_{j,m-j}=O(e^m)$,   where $e$ is either planet's eccentricity (Eqs. \ref{eq:R_zeroth}--\ref{eq:R_second}).
  Evaluating Equation (\ref{eq:TTV_Fourier_AliasedII}) at zeroth-order in $e$ implies
\begin{equation}
	\mu' t^{(0)}_j   =  z_{j,1-j}- z^*_{-j,1+j}-i\lambda_{j,-j} \label{eq:dt0}\nonumber
\end{equation}
  Both zeroth-order and first-order MMR's contribute to this expression: the former through $\lambda_{j,-j}\propto R_{j,-j}$, and the latter through $z_{j,1- j}\propto {\partial R_{j,1- j}/\partial z^*}$.  Inserting the expressions for the $z$'s and $\lambda$'s (Eqs. \ref{eq:zhat}--\ref{eq:lhat})
 and then for the $R$'s (Eqs. \ref{eq:R_zeroth}--\ref{eq:R_second})
  yields
\begin{eqnarray}
	 t^{(0)}_j &=&
		  \frac{
		  2
		  }{\sqrt{\alpha}}\left( 
		             n_{j,1-j}\frac{\partial R_{j,1-j}}{\partial z^*}-n_{-j,1+j}\left(\frac{\partial R_{-j,1+j}}{\partial z^*}\right)^*\right) - 
			\frac{3j}{\alpha^2} n_{j,-j}^2R_{j,-j}+2n_{j,-j} \sqrt{\alpha}\frac{\partial}{\partial \alpha}R_{j,-j} \label{eq:dt0_explicitI}
			\\
			& =&\frac{1}{\sqrt{\alpha}}\left(n_{j,1-j}\left(f_{27}^j+\delta_{1,j}\frac{3}{2}\alpha\right)-n_{-j,j+1} \left( f_{27}^{-j}-\delta_{1,j}\frac{1}{2}\alpha\right)\right) - \left(\frac{3j}{\alpha^2} n_{j,-j}^2-2n_{j,-j}\sqrt{\alpha}\frac{\partial}{\partial \alpha}\right)\left(f_1^j - \delta_{j,1}\frac{\alpha}{2}\right) \label{eq:dt0_explicitII} 
\end{eqnarray}
The quantities entering in this  expression are all roughly of order unity, with the possible exception of  $n_{j,1-j}$,
 which is large  at $j=J$ when the planets lie near a $J$:$J$-1 MMR.
\item {\bf $t_j^{(1)}$:} Following the same reasoning as above, 
	\begin{equation}
		 \mu't^{(1)}_j   =   z_{j,2-j}- z^*_{-j,2+j}-i\lambda_{j,1-j} \label{eq:dt1}
		 \nonumber
	\end{equation}
	 For most values of $j$,  the $t^{(1)}_j$  are  small $O(e)$ corrections to  $t^{(0)}_j$. However,  for a planet pair  near a first-order $J$:$J$-1  MMR,
the factor $n_{j,1-j}$ is large at $j=J$, and that factor can compensate for the smallness of $e$.  Similarly,  for a pair  near a second-order $K$:$K$-2  MMR, the factor $n_{j,2-j}$   is large at $j=K$.	We therefore  approximate $t^{(1)}_j$ by keeping only  terms that are potentially made large by proximity to an MMR:
	 
	\begin{eqnarray}
		t_j^{(1)}\approx t^{(1)}_{j,\cal F}+t^{(1)}_{j,\cal S}
	\end{eqnarray}
		where 
	\begin{eqnarray}
		t^{(1)}_{j,\cal F}&=&\frac{3(1-j)}{\alpha^2}n_{j,1-j}^2 R_{j,1-j}  \label{eq:tF_1}
			\\
		&=&  {3(1-j)\over 2\alpha^2}n_{j,1-j}^2\left(f_{27}^jz^*+f_{31}^jz'^*-2\alpha\delta_{j,2}z'^* \right)
			\\
		t^{(1)}_{j,\cal S}&=& \frac{2 n_{j,2-j}}{\sqrt{\alpha}}\frac{\partial R_{j,2-j}}{\partial z^*} \label{eq:tS_1}
			\\
		&=&  \frac{2 n_{j,2-j}}{2\sqrt{\alpha}}\left(2f_{45}^jz^*+f_{49}^jz'^*   \right)
	\end{eqnarray}
	 where, at the risk of proliferation of subscripts, the ${\cal F}$ component is potentially 
	large near a first order MMR, while the ${\cal S}$ component is potentially large near a second order 
	MMR.	Note also that we  drop
       a term $\propto n_{j,1-j}$,  since it will be much smaller than the  $n_{j,1-j}^2$ in the
       ${\cal F}$ component  when either is important.
	
\item {\bf $t_j^{(2)}$:}
Equation 
 (\ref{eq:TTV_Fourier_AliasedII}) implies
	\begin{equation}
		\mu' t^{(2)}_j   = -i\lambda_{j,2-j}\label{eq:dt2} \ , \nonumber
	\end{equation}where we have ignored the $z$ terms because they can only be large if the planet pair is near a third-order MMR, a possibility we exclude. Again, only terms that are large near MMRs will make a significant contribution to the total TTV. Reasoning as before, 	we  approximate 
	 \begin{eqnarray}
		 t^{(2)}_j  &\approx& \frac{3(2-j)}{\alpha^2}n_{j,2-j}^2 R_{j,2-j} \label{eq:t2j} \\
	&=&\frac{3(2-j)}{2\alpha^2}n_{j,2-j}^2\left( f_{45}^j z^{*2} +f_{49}^jz^*z'^* +\left( f_{53}^j -\delta_{j,3}\frac{27\alpha}{8}\right) z'^{*2}\right) 
	\end{eqnarray}	
	
\end{itemize}

To summarize, the TTV of a planet with an external perturber is given by Equation \eqref{eq:dtAll}, 
with  coefficients $t_j^{(m)}$ as listed in this subsection.  In order to interpret observed TTV's it is helpful to decompose
 the sum in Equation \eqref{eq:dtAll} into  terms with distinct temporal frequencies, as described in 
 \S \ref{sec:analytic_model}, and also to drop all subdominant terms at a given frequency.  We consider the two cases of relevance separately:
 \begin{itemize}
   \item {Companion near $J$:$J$-1 resonance:}
 We decompose the sum as
	$\delta t =  \delta t_{\cal F}+\delta t_{\cal C} + \delta t_{\cal S} $, 
	where the subscripts stand for fundamental, chopping, 
and secondary  (see Eq. \ref{eq:decomp}), where
\begin{eqnarray}	
	\delta t_{\cal F} &=& 
	\mu'{P\over 2\pi i}\left(
	 t^{(0)}_J +  t^{(1)}_{J,{\cal F}} \right) e^{iJ\lambda'_0}  
	+ c.c. \label{eq:tfund} \\
		\delta t_{\cal C} &=& 
	\mu'{P\over 2\pi i}
	\sum_{j>0,j\ne J}  t^{(0)}_j e^{ij\lambda'_0}  + c.c. \ , \label{eq:tchop} \\
	\delta t_{\cal S} &=& 
	\mu'{P\over 2\pi i}
	 \left( t^{(1)}_{2J,{\cal S}} +  t^{(2)}_{2J}\right) e^{2iJ\lambda'_0} 
	 + c.c.  \label{eq:tsec}
\end{eqnarray}
At $O(e^0)$ (i.e., terms with superscript $0$), we
 transfer the $j=J$ term from the sum in  Equation \eqref{eq:tchop} to Equation \eqref{eq:tfund} because it has
 the same frequency and can have comparable amplitude; at $O(e)$, 
we only include the  $j=J$ and $j=2J$ because they 
 are the only ones with  near-resonant denominators; and  similarly at $O(e^2)$ we only include the $j=2J$ term.  Note that the  $\delta t_{\cal F}$ term has the longest period (given by Eq. \ref{eq:super})  because the expressions are  evaluated at the transit times of the inner planet ($\lambda_0=0$).
 
\item {Companion near $K$:$K$-2 resonance, with $K$ odd:} We decompose the sum as 
  $\delta t=\delta t_{\cal C}+\delta t_{\cal S}$ where
     \begin{eqnarray}
         \delta t_{\cal C}&=& \mu' {P\over 2\pi i} \sum_{j>0}  t^{(0)}_j e^{ij\lambda'_0}+c.c. \\
         \delta t_{\cal S}&=& \mu'{P\over 2\pi i}
	 \left( t^{(1)}_{K,{\cal S}} +  t^{(2)}_{K}\right) e^{iK\lambda'_0} + c.c.
	 \label{eq:ts2}
     \end{eqnarray}

 \end{itemize}

\subsection{\label{sec:outerPlanet}\ref{sec:outerPlanet}: Explicit TTV Formulae (Internal Perturber)}

Thus far we have considered the case of an external perturber.  Here we work through the
case of an internal perturber. Since it is largely similar, we skip many of the details.  
The equations of motion (Eqs. \ref{eq:zdot}--\ref{eq:ldot}) become
	\begin{align}
		&\frac{dz'}{dt} = 2i n' \mu \pd{R}{z'^*}   \label{eq:z1dot} \\
		&\frac{d\ln a'}{dt} = 2n' \mu \pd{R}{\lambda'}  \label{eq:a1dot} \\
		&\frac{d\lambda'}{dt} \approx n'\left(1-\frac{3}{2} \frac{\delta a'(t)}{a'}\right) + 2 n' \mu \left( 1 + \alpha \pd{}{\alpha}  \right) R \label{eq:l1dot}
	\end{align}
The disturbing function is the same  as before	(Eqs. \ref{eq:R_Form}--\ref{eq:R_second}), except for the indirect terms:
the coefficients of the Kroenecker delta's are to be replaced by
\begin{eqnarray}
		R_{1,-1} &:~& \frac{\alpha}{2} \longrightarrow \frac{1}{2\alpha^2}\\
		R_{2,-1} &:~& 2\alpha \longrightarrow \frac{1}{2\alpha^2}	 \\
		R_{3,-1} &:~& \frac{27\alpha}{8}\longrightarrow \frac{3}{8\alpha^2}
\end{eqnarray}
 The expansion in eccentricity (Eq. \ref{eq:dtAll}) becomes
 \begin{equation}
 \delta t'= \mu\frac{P'}{2 \pi i}
 \sum_{j<0}
 \left(
 	t'^{(0)}_j+t'^{(1)}_j + t'^{(2)}_j
  \right)e^{ij\lambda_0} + c.c. \label{eq:dt_outer} \ .
\end{equation}
Note that we choose here the sum to be over negative $j$'s as this allows the $t'^{(m)}_j$ to be expressed in terms of the  $R_{j,k}$ listed in Equations \eqref{eq:R_zeroth}---\eqref{eq:R_second} (a sum over positive $j$ values would require the complex conjugates, $R^*_{j,k}$).   
   
The coefficients are: 
\begin{eqnarray}
\label{eq:dt10_explicit}
	 t'^{(0)}_{-j}	&=& n_{1+j,-j} f_{31}^{(j+1)} -n_{1-j,j} f_{31}^{(1-j)}+ \left(3 j  n_{j,-j}^2 -2 n_{j,-j}\left(1+\alpha\frac{\partial }{\partial\alpha}\right)\right)\left(f_1^j - \delta_{j,1}\frac{1}{2\alpha^2}\right) \\
	 t'^{(1)}_{-j}	&\approx&  \left
	(n_{j+2,-j}\left( f_{49}^{(j+2)}z^* + 2   f_{53}^{(j+2)}z'^*- \delta_{j,1}\frac{3}{4\alpha^2}z'^*\right) + 
	\frac{3(j+1)}{2}   n_{j+1,-j}^2 \left( f_{27}^{(j+1)} z^*+f_{31}^{(j+1)} z'^*- \delta_{j,1}\frac{1}{2\alpha^2}z'^*  \right)
	\right)\label{eq:dt11_full}
\\
	 t'^{(2)}_{-j}	&\approx&  \left( \frac{3(j+2)}{2}  n_{j+2,-j}^2  \left( f_{45}^{(j+2)} {z^*}^2 +f_{49}^{(j+2)} z^* z'^* +f_{53}^{(j+2)}{z'^*}^2- \delta_{j,1}\frac{3}{8\alpha^2}z'^{*2} \right) \right) \label{eq:dt12_full}
\end{eqnarray}

Finally, the decomposition into terms with distinct temporal frequencies is essentially the same as 
Equations  \eqref{eq:tfund}--\eqref{eq:ts2}, and after the appropriate replacements:
\begin{eqnarray}	
	\delta t'_{\cal F} &=& 
	\mu{P'\over 2\pi i}\left(
	 t'^{(0)}_{1-J} +  t'^{(1)}_{1-J,{\cal F}} \right) e^{i(1-J)\lambda_0}  
	+ c.c. \label{eq:t1fund} \\
		\delta t'_{\cal C} &=& 
	\mu{P'\over 2\pi i}
	\sum_{j<0,j\ne1- J}  t'^{(0)}_j e^{ij\lambda_0}  + c.c. \ , \label{eq:t1chop} \\
	\delta t'_{\cal S} &=& 
	\mu{P'\over 2\pi i}
	 \left( t'^{(1)}_{2-K,{\cal S}} +  t'^{(2)}_{2-K}\right) e^{i(2-K)\lambda_0} 
	 + c.c.  \label{eq:t1sec}
\end{eqnarray}
where $J$ and $K$ still refer to the nearest $J$:$J$-1 or $K$:$K$-2 resonance (for $J,K>0$,  and for the case of a first-order MMR, $K=2J$ ).

\subsection{\label{sec:zsection}\ref{sec:zsection}: Simplified Dependence on ${\cal Z}$}
Here we reparameterize $\delta t_{\cal F}$ and $\delta t_{\cal S}$, which have a rather unweildy dependence on $z$ and $z'$, in terms of the single variable ${\cal Z}$ introduced in Equation \eqref{eq:Z_def} by exploiting some approximate relationships between the `$f$' coefficients appearing in the TTV formulae.
We carry out the derivation for a planet with an exterior companion; the derivation for planets with interior companion is completely analogous and we merely quote the final result.
We assume that the planet is not near a 2:1 or 3:1 MMR since the TTV formulae near these MMRs are complicated by the contribution of indirect terms (see Section \ref{sec:inferring}).
Using the definition of $\cal Z$ from Equation \eqref{eq:Z_def}, the eccentricity-dependent component of the fundamental TTV,  $\delta t^{(1)}_{J,\cal F}$ (Eq. \ref{eq:tF_1}), can trivially be rewritten as 
\begin{equation}
 t^{(1)}_{J,\cal F} = \frac{3(1-J)}{2\alpha^2}n_{J,1-J}^2 \sqrt{(f^J_{27})^2+(f^J_{31})^2} {\cal Z}^* \label{eq:dt1_mmr_1}  \ . 
 \end{equation}

Next we reparameterize $\delta t_{\cal S}$ in terms of ${\cal Z}$. We first consider $\delta t_{\cal S}$ near a first-order $J$:$J$-1 MMR; the extension to second-order $K$:$K$-2 MMRs, described below, is trivial.
The first step in simplifying $\delta t_{\cal S}$  is rewriting  $R_{J,2-2J}$ as:
\begin{align}
	R_{J,2-2J} &= \frac{1}{2}\left(f^{2J}_{45}z^{*2} +f^{2J}_{49}z^{*}z'^{*}+f^{2K}_{53}z'^{*2}\right)\approx \frac{1}{2}\gamma {\cal Z}^{*2}  \label{eq:fRelation} \\
	\gamma &\equiv \frac{f^{2J}_{49}}{2f^{J}_{27}f^{J}_{31}} \left( (f_{27}^{J})^2+ (f_{31}^{J})^2 \right) \label{eq:gamma_def}  \ .
\end{align}
 Equations \eqref{eq:fRelation} and  \eqref{eq:gamma_def} warrant a few remarks. 
First, the approximation in Eq. \eqref{eq:fRelation} expresses apparently coincidental relationships between Laplace coefficients, namely:
$ f_{45}^{2J}/(f_{27}^{J})^2 \approx f_{53}^{2J}/(2f_{31}^{J}f_{27}^{J}) \approx f_{53}^{2J}/(f_{31}^{J})^2$.
Thus, the coefficients of each the quadratic terms in $z$ and $z'$ are equal or nearly equal in the left- and right-hand side of Equation \eqref{eq:fRelation}.
Equation \eqref{eq:fRelation} is extended to second-order $K$:$K$-2 MMRs by replacing 2$J$ with $K$ and defining ${\cal Z}$ in terms of $f_{27}^J$ and
 $f_{31}^J$ ( Eq. \ref{eq:Z_def}) by taking $J=\lceil K/2\rceil$, that is, $K/2$ rounded up to the nearest whole integer.
The approximation matches the values of $f^{K}_{45}$ and $f^{K}_{53}$ with $<2\%$ fractional error for $5\le K\le 11$ and $|\Delta|<0.02$. 
Substituting Equation \eqref{eq:fRelation} in Equations  \eqref{eq:tS_1} and \eqref{eq:t2j},  $t^{(1)}_{K,{\cal S}}$  and  $t^{(2)}_{K}$ become
\begin{align}
	t^{(1)}_{K,{\cal S}} &=\frac{2n_{K,2-K}}{\sqrt{\alpha}} \frac{\gamma f_{27}^J}{\sqrt{(f_{27}^{J})^2+ (f_{31}^{J})^2}} {\cal Z}^* \label{eq:dt1_mmr_2}\\
	t^{(2)}_{K} &= \frac{3(2-K)}{2\alpha^2}n_{K,2-K}^2  \gamma{\cal Z}^{*2} 	\label{eq:dt2_mmr_2} \ .
 \end{align}

In Equations \eqref{eq:tfund}--\eqref{eq:tsec} we account for the eccentricity-dependent TTV contributions of only the {\it nearest} first and/or second MMRs, which we have parameterized in terms of ${\cal Z}$. 
In fact,  to good approximation, the contributions of 
all\footnote{This excludes contributions of the 2:1 and 3:1 MMRs to the TTV because of the associated indirect terms. 
Planets near any other MMR will be far away from the 3:1 and 2:1 resonances and so the $O(e)$ and $O(e^2)$ contributions of these MMRs to the total TTV will be small.} 
first- and second-order MMRs  depend on the planets' complex eccentricities only through the single combination, ${\cal Z}$.
Additional eccentricity-dependent terms are increasingly important as the planet period ratio approaches unity and successive first- and second-order MMRs become more closely spaced.
The TTV formulas can be generalized to incorporate the effects of additional  first- and second-order MMRs by adding the appropriate $t_{j,\cal F}^{(1)}$, $ t_{j,\cal S}^{(1)}$ and $t_{j}^{(2)}$ terms, defined by Equations \eqref{eq:tF_1},\eqref{eq:tS_1}, and \eqref{eq:t2j},  to the formulas.
The additional terms can be expressed in terms of ${\cal Z}$ using Equations \eqref{eq:dt1_mmr_1}, \eqref{eq:dt1_mmr_2}, and \eqref{eq:dt2_mmr_2}, by simply replacing $J$ and $K$  (though, importantly, not in the definition of ${\cal Z}$) with the appropriate integer.
This is because ratio  of the `$f$' coefficients that determines $\cal Z$ , i.e. $f^j_{27}/f^j_{31}$, is nearly independent of the integer  $j$ and is instead primarily determined by the period ratio of the planets (the ratio of $f^j_{27}/f^j_{31}$ varies with $j$ by less than $3\%$ for $3\le j\le6$ when evaluated at a fixed period ratio in the range ${ 2 / 3}\le{P / P'}\le{ 5 / 6}$).
The combination  of $z$ and $z'$  that appear in the contribution of a particular MMR to the TTV is determined mainly by the planets' period ratio and depends only weakly on the particular MMR, allowing 
  Equations \eqref{eq:dt1_mmr_1}, \eqref{eq:dt1_mmr_2}, and \eqref{eq:dt2_mmr_2} to be used to approximate TTV contribution of any and all nearby first- and second-order MMRs.

Inserting the definition of $\cal Z$ and Equation \eqref{eq:fRelation} into Equations \eqref{eq:dt11_full} and \eqref{eq:dt12_full}, the components comprising the fundamental and secondary TTV of a planet with an interior perturber become
\begin{align}
	 t'^{(1)}_{1-J,\cal F} &= n_{J,1-J}^2 \frac{3J}{2}   \sqrt{(f^J_{27})^2+(f^J_{31})^2}   {\cal Z}^*  \label{eq:dt11_mmr_1}\\
	t'^{(1)}_{2-K,{\cal S}} &= 2n_{K,2-K} \frac{\gamma f_{31}^J}{\sqrt{(f_{27}^{J})^2+ (f_{31}^{J})^2}} {\cal Z}^* \label{eq:dt11_mmr_2}\\
	t'^{(2)}_{2-K} &=  \frac{3K}{2}  n_{K,2-K}^2 \gamma {\cal Z}^{*2} \label{eq:dt12_mmr_2}\ .
 \end{align}
Numerical values for the coefficients appearing in Equations \eqref{eq:dt1_mmr_1}, \eqref{eq:dt1_mmr_2}, and \eqref{eq:dt2_mmr_2} and Equations \eqref{eq:dt11_mmr_1}--\eqref{eq:dt12_mmr_2} are listed in Table \ref{tab:coeffs}. 

\begin{table*}
\caption{TTV Coefficients}
\begin{center}
\begin{tabular}{| l | l l l l l l  | l l l  |}
\hline
Nearest Resonance &$ t^{(0)}_{1}$ & $ t^{(0)}_{2}$ & $ t^{(0)}_{3}$ & $ t^{(0)}_{4}$ & $ t^{(0)}_{5}$ & $ t^{(0)}_{6}$ & $ t^{(1)}_{J,{\cal F}}/{\cal Z}^*$ & $ t^{(1)}_{K,{\cal S}}/{\cal Z}^*$ & $ t^{(2)}_{K}/{\cal Z}^{*2}$ \\ \hline
3:2~($J=3$)&-6.5	&-10.4	&-2.8+$0.8\over\Delta$	&2.5	&0.7	&0.3				&-1.8$\Delta^{-2}$	&3.3$\Delta^{-1}$	&-3.9$\Delta^{-2}$\\
7:5~($K=7$)&-10.7	&-13.5	&-18.7	&10.2		&2.1	&0.8					& ---	&3.9$\Delta^{-1}$	&-4.6$\Delta^{-2}$\\
4:3~($J=4$)&-16.0	&-17.6	&-16.7	&-4.2+$0.8\over\Delta$	&5.8	&1.9			&-1.8$\Delta^{-2}$	&4.6$\Delta^{-1}$&-5.3$\Delta^{-2}$\\
9:7~($K=9$)&-22.6	&-22.7	&-18.3	&-31.4	&19.8	&4.4					&---	&5.3$\Delta^{-1}$	&-6.0$\Delta^{-2}$\\
5:4~($J=5$)&-30.6&-28.7&-21.2&-24.6&-5.6+$0.8\over\Delta$&10.7					&-1.8$\Delta^{-2}$	&6.0$\Delta^{-1}$&-6.7$\Delta^{-2}$\\
\hline \hline
Nearest Resonance & $ t'^{(0)}_{-1}$ & $ t'^{(0)}_{-2}$ & $ t'^{(0)}_{-3}$ & $ t'^{(0)}_{-4}$ & $ t'^{(0)}_{-5}$ & $ t'^{(0)}_{-6}$ & $ t'^{(1)}_{1-J,{\cal F}}/{\cal Z}^*$ & $ t'^{(1)}_{2-K,{\cal S}}/{\cal Z}^*$ & $ t'^{(2)}_{2-K}/{\cal Z}^{*2}$\\ \hline
3:2~($J=3$)&6.8	&4.3-$0.8 \over \Delta$	&-2.2	&-0.6	&-0.2	&-0.1	&1.6$\Delta^{-2}$	&-3.5$\Delta^{-1}$	&3.4$\Delta^{-2}$\\
7:5~($K=7$)&10.4	&21.9	&-10.1	&-1.9	&-0.7	&-0.3			&---	&-4.2$\Delta^{-1}$	&4.1$\Delta^{-2}$\\
4:3~($J=4$)&15.1	&20.9	&5.8-$0.8\over\Delta$	&-5.5	&-1.7	&-0.8	&1.6$\Delta^{-2}$	&-4.9$\Delta^{-1}$	&4.8$\Delta^{-2}$\\
9:7~($K=9$)&21.2	&24.0	&34.6	&-19.8	&-4.2	&-1.7			&---	&-5.6$\Delta^{-1}$	&5.5 $\Delta^{-2}$\\
5:4~($J=5$)&28.5	&28.8	&28.2	&7.2-$0.8\over\Delta$	&-10.5	&-3.6	&1.6$\Delta^{-2}$	&-6.2$\Delta^{-1}$	&6.3$\Delta^{-2}$\\ \hline
\end{tabular}
\end{center}
\tablecomments{
Numerical values of various components of the analytic TTV  formulae appearing in Equations \eqref{eq:dt0_explicitII}, \eqref{eq:dt1_mmr_1}, \eqref{eq:dt1_mmr_2}, and \eqref{eq:dt2_mmr_2}.
Each row of the table lists numerical values for components near a particular first- or second-order MMR.  
The numerical values are computed at the location of exact resonance. Most of the coefficients depend weakly on planet period ratio: 
a median fractional error of $\sim8|\Delta|$ is incurred though in some instances fractional errors can be in excess of $50|\Delta|$.
}
\label{tab:coeffs}
\end{table*}%

 \subsection{\label{sec:inclinations}\ref{sec:inclinations}: Mutual Inclinations}
Here we briefly consider the influence of mutual inclinations on the TTV signal.
Inclinations, $I$, only enter $R_{j,-j}$ and $R_{j,1-j}$  through terms of order $e I^2$ and higher so that 
$\delta t_{\cal C}$ and $\delta t_{\cal F}$ are essentially unchanged for moderate values of mutual inclination.
 We need only consider the contributions of inclinations to second-order MMRs in our TTV formulae.
Mutual inclinations introduce an additional term to the disturbing coefficient $R_{j,2-j}$ (Eq. \ref{eq:R_second}) given by:  
 \begin{align}
 	 R^\text{(inc)}_{j,2-j} &= \frac{1}{2}f^j_{57} \xi^{*2} \\
 	\xi &\equiv \sin(I/2)\exp(i\Omega) - \sin(I'/2)\exp(i\Omega')
 \end{align}
 where $\Omega$ is the longitude of ascending node. 
 Incorporating this term into the TTV formulae near second order resonances is straightforward and Equations \eqref{eq:dt2_mmr_2} and \eqref{eq:dt12_mmr_2} for the secondary TTV signals become:
 \begin{eqnarray}
 t^{(2)}_{K} &= \frac{3(2-K)}{2\alpha^2}n_{K,2-K}^2  \left(\gamma {\cal Z^*}^2 + f^K_{57}\xi^{*2} \right) \\
 t'^{(2)}_{2-K}  &= \frac{3K}{2} n_{K,2-K}^2 \left(\gamma {\cal Z^*}^2 + f^K_{57}\xi^{*2} \right) 
 \end{eqnarray}
We ignore the contribution of mutual inclinations to the secondary TTV because for $|\xi |\sim |{\cal Z}|$ their contribution to $t_{\cal S}$ will be small since $f^K_{57}/\gamma  < 0.2$ (for $5\le K\le11$).  


\section{B:   MCMC Methods}
\subsection{\label{sec:nbody_mcmc} \ref{sec:nbody_mcmc}: MCMC with N-body}
	We model each planetary system as point masses orbiting a central star and compute mid-transit times via N-body integration. We use the \verb|TTVFast| code developed by \citet{2014ApJ...787..132D} to compute transit times. Planets are assumed to have coplanar orbits. We carry out Markov Chain Monte Carlo (MCMC) analyses of each system to infer planet masses and orbits. The MCMC analyses of each multi-planet system are carried out using the \verb|EMCEE| package's \citep{ForemanMackey:2012io} ensemble sampler. The \verb|EMCEE| package employs the algorithm of \citet{Goodman:2010eta} to evolve an ensemble of `walkers' in parameter space, with each walker yielding a separate Markov chain of samples from the posterior distribution. 
    
    The parameters of the MCMC fits are each planet's planet-to-star mass ratio, $\mu_i$, eccentricity vector components $h_i \equiv e_i\cos(\varpi_i)$ and $k_i \equiv e_i\sin(\varpi_i)$, initial osculating period, $P_i$, and initial time of 
transit\footnote{In reality, we use the parameters, $T_i$, as a convenient re-parameterization of the planets mean longitudes, $\lambda_i$, so that  $\lambda_i=(\text{epoch}-T_i)/2\pi$ for a chosen reference epoch.}
, $T_{i}$ where $i=1,2,...,N$ and $N$ is the number of planets. Errors in the observed transit times are assumed to be independent and Gaussian with standard deviations given by the reported observational uncertainty so that the likelihood of any set of parameters is proportional to $\exp(-\chi^2/2)$, where $\chi^2$ has the standard definition in terms of normalized, squared residuals:
\begin{equation}
\chi^2 = \sum_{i=1}^N \sum_{j:\text{transits}} \left(\frac{t_{\text{obs.},i}(j) - t_{\text{N-body},i}(j)}{\sigma_{i}(j)}\right)^2
\end{equation}
 where the $t_{\text{obs.},i}(j)$ are the observed transit times, indexed by $j$, of the $i$th planet, $\sigma_i(j)$ are their reported observational uncertainties, and $t_{\text{N-body},i}(j)$ are the transit times computed by N-body integration. We begin each MCMC ensemble by searching parameter space for a minimum in $\chi^2$ with a Levenberg-Marquardt (LM) least-squares minimization algorithm \citep[e.g.,][]{press1992nr}. Transit time observations that fall more than 4-$\sigma$ away from the initial best fit, measured in terms of the reported uncertainty, are marked as outliers and removed from the data. We find that our MCMC results are largely insensitive to the removal of outliers, having experimented with fitting transit times with outliers included as well as more liberally removing poorly fit transit times. The new transit times are then refit with the LM algorithm and an ensemble of walkers are initialized  in a tight `ball' around the identified minimum. This is done by drawing the walkers' initial positions from a multivariate Gaussian distribution based on the estimated covariance matrix generated by the LM algorithm.

  We estimate the number of independent posterior samples generated by each MCMC run based on the auto-correlation length of each walker's Markov chain. This is done as follows. First, for each walker in an ensemble, we compute the auto-correlation functions,
  \begin{equation}
\rho_i(\tau) = \frac{<X_i(s)X_i(s+\tau)>-<X_i(s)>^2}{<X_i(s)X_i(s)>-<X_i(s)>^2} 
\end{equation}
 where $<...>$ denotes the average over sample number, $s$, and the $X_i$ denote the various model parameters, with $i$ ranging from $i=1,..,5N$ for a system of $N$ planets.  We then take the auto-correlation length in each parameter to be the value of $\tau$ at which $\rho_i(\tau)$ decreases one $e$-folding, i.e.,  $\rho_i(\tau)<e^{-1}\approx 0.37$.  We assign an auto-correlation length to each walker that is the maximum auto-correlation length, over all the $5N$ model parameters, in that walker's Markov chain. Finally, the number of independent posterior samples generated by an individual walker during an MCMC run is taken to be the total number of samples in the chain divided by the walker's auto-correlation length. The full posterior samples generated by each MCMC fit are available online at \url{https://sites.google.com/a/u.northwestern.edu/shadden}.

For each system presented in Section \ref{sec:systems} we ran MCMC simulations with two different priors: default and `high mass'.
Both priors are uniform in all planets' periods, $P_i$, and times of initial transit $T_i$. 
Furthermore, we assume the prior probabilities of each planets' masses and eccentricities are independent.  
Therefore the prior probability density for a set of MCMC parameters, ${\theta}$, of an $N$-planet system can be written as
\begin{equation}
\text{Prob}({\theta})d{\theta} = \prod_{i=1}^N p(\mu_i)p(h_i,k_i) {d\mu_i}{dh_i}{dk_i}{dT_i}{dP_i}
\end{equation}
where $p(\mu_i)$ and $p(h_i,k_i)$ are the marginal prior probabilities in a planet's mass and eccentricity components, respectively. The prior probability density, $p(h_i,k_i)$, for a planet's eccentricity components can be expressed in terms of the planet's eccentricity, $e_i$,  and longitude of periapse, $\varpi_i$, as \citep{2006ApJ...642..505F}:
\begin{equation}
	p(h_i,k_i)dh_i dk_i = p(e_i\cos\varpi_i,e_i\sin\varpi_i) e_i de_i d\varpi_i \label{eq:probability_transform}
\end{equation}
where the factor of $e_i$ arises from the Jacobian of the coordinate transformation $(h_i,k_i)\rightarrow(e_i,\varpi)$. 
Both the default and high mass prior probability densities have the functional forms:
\begin{align}
p(\mu) \propto \begin{cases} 
					\mu^{-\alpha} &;~\mu \ge 0 \\
                      0	&;	\text{ otherwise}	
                    \end{cases} \label{eq:mu_prior}\\
p(h,k) \propto \begin{cases} 
					(h^2+k^2)^{-\beta/2} &;~(h^2+k^2)^{1/2} < 0.9 \\
                      0					&;\text{ otherwise}	
                    \end{cases} \label{eq:eprior}
\end{align}
each with a different value for the exponents $\alpha$ and $\beta$. We impose the condition $(h^2+k^2)^{1/2} < 0.9$ to avoid evaluating N-body integrations that require exceptionally small time steps.
 In practice, we find that the posterior probability densities are negligible at eccentricities well below this imposed upper bound, thus it does not influence our conclusions. 
 For our default priors we set $\alpha=1$ and $\beta = 1$ in Equations \eqref{eq:mu_prior}  and \eqref{eq:eprior}.
 The choice of $\alpha=1$ yields a prior that is uniform in $\log(\mu)$. 
 This choice is typical as a non-informative prior for positive-definite ``scale" parameters \citep{gregory2005bayesian}. 
Setting $\beta=1$ results in a prior that is uniform in eccentricity since inserting Equation \eqref{eq:eprior} into  \eqref{eq:probability_transform} gives $p(h,k)dhdk\propto e^{-\beta+1}ded\varpi$.
For the high mass priors we set  $\alpha=0$ and $\beta = 2$. The resulting priors are uniform in $\mu_i$ and $\log(e_i)$.
This combination favors more massive planets as explained in Section \ref{sec:methods}.

\subsection{\label{sec:analytic_mcmc} \ref{sec:analytic_mcmc}: MCMC with Analytic Model}

 We also carry out full MCMC analyses of each TTV system using the analytic model.
 In two planet systems, the TTVs of both planets are fit as a function of the planet-to-star mass ratios and the combined eccentricity, $\cal Z$. 
 We only include the pairwise interactions of adjacent planets when fitting the four planets of the Kepler-33 system (Section \ref{sec:kep33}). 
 The analytic formulas give TTVs as a function of the planet-to-star mass ratios and the combined complex eccentricity, ${\cal Z}$.
 This constitutes a significant reduction in the number of required model parameters required for TTV fitting: from the $5 \times N$ parameters, where $N$ is the number of planets, required for a coplanar N-body fit (see Section \ref{sec:nbody_mcmc}), to the parameters of the analytic model: one planet-star mass ratio for each planet considered and two components of ${\cal Z}$ for each pairwise interaction considered. 
 
To carry out MCMC fits with the analytic model each planet's transit times are first converted to TTVs.
Converting   transit times to TTVs requires determining a planet's average period.
Average periods are determined by fitting the transit times of planets near first-order MMRs as the sum of a linear trend plus sinusoidal terms with the frequencies expected for the principal and secondary TTV components.
If a planet pair is near a second order MMR then their transit times are fit as the sum of a linear trend plus the secondary TTV component.
Since the frequencies of the principal and secondary TTV signals depend on the planet periods, we fit the transit times of all planets in a system simultaneously with a nonlinear LM fit. \
The best-fitting linear trends are subtracted from the observed transit times to yield the TTVs fit by the MCMC.

The likelihood of a set of parameters in the analytic MCMC is computed from their $\chi^2$ value as in the N-body MCMC. The TTV of the inner planet is computed in the analytic MCMC according to Equation \eqref{eq:dtAll} by including $t_j^{(2)}$ only for $j=K$ where $K$:$K$-2 is the nearest second-order MMR (including $K$=$2J$ near a $J$:$J$-1 MMR) and including all $t_j^{(0)}$ and $t_j^{(1)}$ terms for $1\le j \le K$. The term $t_K^{(2)}$, as well as each $t_j^{(1)}$, is a function of the variable ${\cal Z}$ and is computed according to the approximations discussed in Appendix \ref{sec:zsection}. The TTV of the outer planet is computed similarly using Equation \eqref{eq:dt_outer} with the terms $t_{2-K}'^{(2)}$ and $t_j'^{(0)}$ and $t_j'^{(1)}$ for $1\le j \le 2- K$ included.

 MCMC analyses using the analytic models are carried out using the {\it Kombine} MCMC code\footnote{\url{http://home.uchicago.edu/~farr/kombine}} { (Farr \& Farr, in prep)}.
 {\it Kombine} is an ensemble sampler that iteratively constructs a kernel-density-estimate-based proposal distribution to approximate the target posterior distribution.
 With {\it Kombine}, the proposal distribution is identical for each Markov chain in the ensemble and is computed to approximate the underlying posterior distribution
 which allows independent samples to be more rapidly generated  than \verb|EMCEE|.
 We find that the {\it Kombine} code fails to converge to a proposal distribution with a high acceptance fraction when using the N-body TTV model.
 Our analytic MCMC uses priors that are uniform in $\log(\mu_i)$ and $|{\cal Z}|$. 

\end{document}